\documentclass[conference]{IEEEtran}
\IEEEoverridecommandlockouts
\usepackage{cite}
\usepackage{amsmath,amssymb,amsfonts}
\usepackage{algorithmic}
\usepackage{graphicx}
\usepackage{textcomp}
\usepackage{xcolor}
\usepackage[caption=false]{subfig} 
\usepackage{booktabs}
\usepackage{multirow}
\usepackage{pifont}
\usepackage{xurl}
\usepackage{hyperref}
\usepackage{listings}

\def\BibTeX{{\rm B\kern-.05em{\sc i\kern-.025em b}\kern-.08em
    T\kern-.1667em\lower.7ex\hbox{E}\kern-.125emX}}

\usepackage[most]{tcolorbox} 
\usepackage{enumitem}        
\newcommand{\llmplaceholder}[1]{\texttt{\{#1\}}}
\newtcolorbox{llmscriptblock}[1][]{
    breakable,
    colback=black!5!white,      
    colframe=black!60!white,    
    fonttitle=\bfseries\sffamily, 
    coltitle=black,
    sharp corners,
    left=8pt,  
    right=8pt, 
    top=5pt,    
    bottom=5pt, 
    attach boxed title to top left={yshift=-0.15\baselineskip, xshift=5pt}, 
    boxed title style={
        colback=white,
        colframe=black!60!white,
        sharp corners,
        boxrule=0.5pt,
        arc=0pt,
        bottomrule=0pt, 
        toprule=0pt,   
        rightrule=0pt,
        leftrule=0pt,
        boxsep=2pt, 
    },
    before skip=\medskipamount, 
    after skip=\medskipamount,  
    #1 
}

\newcommand{\cmark}{\ding{51}}
\newcommand{\xmark}{\ding{55}}

\definecolor{lightpurple}{HTML}{D8BFD8}
\definecolor{lightorange}{HTML}{FFDAB9} 
\definecolor{lightcyan}{HTML}{8AE2D8}
\definecolor{lightblue}{HTML}{8AADE2}

\lstdefinelanguage{json}{
    basicstyle=\small\ttfamily,
    stepnumber=1,
    numbersep=8pt,
    showstringspaces=false,
    breaklines=true,
    frame=none,
    literate=
     *{0}{{{\color{blue}0}}}{1}
      {1}{{{\color{blue}1}}}{1}
      {2}{{{\color{blue}2}}}{1}
      {3}{{{\color{blue}3}}}{1}
      {4}{{{\color{blue}4}}}{1}
      {5}{{{\color{blue}5}}}{1}
      {6}{{{\color{blue}6}}}{1}
      {7}{{{\color{blue}7}}}{1}
      {8}{{{\color{blue}8}}}{1}
      {9}{{{\color{blue}9}}}{1}
      {:}{{{\color{black}:}}}{1}
      {,}{{{\color{black},}}}{1}
      {\{}{{{\color{orange}\{}}}{1}
      {\}}{{{\color{orange}\}}}}{1}
      {[}{{{\color{orange}[}}}{1}
      {]}{{{\color{orange}]}}}{1},
}
\definecolor{mdCodeColor}{rgb}{0.85, 0.1, 0.1}  
\definecolor{mdBoldColor}{rgb}{0.3, 0.3, 0.3}  
\definecolor{mdItalicColor}{rgb}{0.1, 0.6, 0.1}
\definecolor{mdQuoteColor}{rgb}{1, 0.5, 0.2} 
\definecolor{mdKeywordColor}{rgb}{0.2,0.2,0.8} 
\definecolor{mdCommentColor}{rgb}{0.5,0.5,0.5} 
\definecolor{mdBraceColor}{rgb}{0.1, 0.6, 0.1} 

\lstdefinelanguage{markdown}{
    basicstyle=\small\ttfamily, 
    sensitive=false,            
    morekeywords={\#, -, *},     
    keywordstyle=\color{mdKeywordColor}\bfseries, 
    moredelim=*[s][\bfseries\color{mdBoldColor}]{**}{**},
    moredelim=*[s][\color{mdKeywordColor}]{\`}{\`},
    moredelim=*[s][\color{mdBraceColor}]{\{}{\}}, 
    morestring=[b]",
    stringstyle=\color{mdQuoteColor}, 
    morecomment=[l]{>},          
    commentstyle=\color{mdCommentColor}\itshape, 
    breaklines=true,           
    keepspaces=true,           
    showstringspaces=false,    
}

\newcommand{\papername}{\textsc{ScaffoldUI}}

\begin{document}

\title{Designing Scaffolded Interfaces for Enhanced Learning and Performance in Professional Software

\author{
    \IEEEauthorblockN{Yimeng Liu}
    \IEEEauthorblockA{
    \textit{University of California, Santa Barbara} \\
    Santa Barbara, USA \\
    yimengliu@ucsb.edu}
    \and
    \IEEEauthorblockN{Misha Sra}
    \IEEEauthorblockA{
    \textit{University of California, Santa Barbara} \\
    Santa Barbara, USA \\
    sra@ucsb.edu}
    }
}

\maketitle

\thispagestyle{plain}
\pagestyle{plain}

\begin{abstract}
Professional software offers immense power but also presents significant learning challenges. Its complex interfaces, as well as insufficient built-in structured guidance and unfamiliar terminology, often make newcomers struggle with task completion. To address these challenges, we introduce \papername{}, a method for scaffolded interface design to reduce interface complexity, provide structured guidance, and enhance software learnability. The scaffolded interface presents task-relevant tools, progressively discloses tool complexity, and organizes tools based on domain concepts, aiming to assist task performance and software learning. To evaluate the feasibility of our interface design method, we present a technical pipeline for scaffolded interface implementation in professional 3D software, i.e., Blender, and conduct user studies with beginners (N=32) and experts (N=8). Study results demonstrate that our scaffolded interfaces significantly reduce perceived task load caused by interface complexity, support task performance through structured guidance, and augment learning by clearly connecting concepts and tools within the taskflow context. Based on a discussion of the user study findings, we offer insights for future research on designing scaffolded interfaces to support instruction, productivity, creativity, and cross-software workflows.
\end{abstract}


\section{Introduction}
Professional software offers significant power and forms the backbone of many creative and design industries. However, this power comes at the cost of significant learning complexity. Prior work has identified that one of the major complexities comes from the user interfaces (UIs), where newcomers often encounter hundreds of icons, menus, and specialized tools~\cite{chilana2018supporting}. They frequently struggle to find the appropriate functions and understand domain-specific terminology~\cite{novick2009micro, kiani2019beyond, lee2010usability}. Compounding these issues is the lack of built-in and structured guidance. This forces users to rely heavily on external resources, such as video tutorials, forums, and more recently, large language models (LLMs), to navigate the interfaces and learn the specific task-related tools, which may disrupt workflow and not always result in contextually relevant help, hindering self-directed learning and exploration~\cite{rieman1996field, chilana2012lemonaid, chilana2018supporting, khurana2024and}.

Multiple approaches have attempted to mitigate these interface-associated challenges. One common strategy has been to provide simplified alternatives, such as dedicated beginner modes~\cite{mcgrenere2002evaluation}, separate novice-oriented applications~\cite{tinkercad}, or even AI-generated task-specific interfaces~\cite{vaithilingam2024dynavis}. While these alternatives can reduce initial complexity, they often embody a simplicity-power tradeoff~\cite{uiandvis1999hearst}. Users may plateau in simplified environments~\cite{cockburn2014supporting} or face a context switch when transitioning to the professional suite, negating prior learning~\cite{nicol2024psychology, li2022design}. 
Another line of work has focused on improving in-context assistance. Efforts range from static tooltips~\cite{grossman2010toolclips} and context-sensitive help~\cite{kelleher2005stencils} to more sophisticated techniques like intelligent user assistance~\cite{horvitz2013lumiere} and systems that embed relevant external resources directly into the application context~\cite{zhong2021helpviz}. However, these approaches can struggle with accurately inferring user goals, potentially leading to intrusive or irrelevant help~\cite{brdnik2022intelligent}, and users may still find it challenging to map the provided information onto the complex software interface~\cite{kosch2023survey}. 
More recently, the advancement of LLMs has led to the emergence of in-application software copilots that can process natural language queries and even automate tasks~\cite{firefly, khurana2025me}. While powerful for well-defined operations, current copilots often struggle with complex, creative, or exploratory tasks~\cite{khurana2024and, qiao2024benchmarking}. Furthermore, excessive automation can restrain the user's ability to learn underlying concepts and workflows~\cite{chen2023understanding, vorvoreanu2025fostering}, diminishing agency~\cite{li2024user, sellen2024rise} and contradicting the well-established preference for learning by doing~\cite{carroll1984training, rieman1996field}.

Despite prior efforts, a significant research gap persists: existing solutions often fail to holistically address UI complexity, provide in-context structured guidance, and support terminology understanding in a way that smoothly integrates learning through task completion into the application workflow. Motivated by instructional scaffolding research where temporary support is provided to help a learner develop skills and ultimately work independently~\cite{belland2017instructional}, we introduce \papername{}, \textbf{a method for designing scaffolded interfaces in professional software} to fill this gap. Our design method aims to address UI complexity, provide structured guidance through UI configuration, and connect UI elements to domain concepts to facilitate learning. Specifically, we focus on four design goals: 

\begin{itemize}
    \item Select and surface task-relevant functionality to reduce UI complexity caused by numerous hidden functions and tools; 
    \item Scaffold learning by disclosing tool complexity as users make progress; 
    \item Organize functionality according to workflow stages and domain concepts to clarify workflow structure and enhance concept learning; 
    \item Promote native software fluency by linking tools presented in the scaffolded interface to the underlying software. 
\end{itemize}

To investigate our method’s feasibility, we presented a technical pipeline to implement our design goals and used it to create two scaffolded interfaces in Blender~\cite{blender}, a professional open-source 3D software, with one interface tailored for a 3D modeling task and the other for an animation task. We evaluated these interfaces with beginner users (N=32) and expert users (N=8). Results indicate that our interfaces significantly improved workflow clarity and task performance for both groups, as well as operation and concept learning for beginners. Based on a discussion of these results, we present insights for future research on using our method to create scaffolded interfaces in professional software for instructional use cases, personalized workflows, and cross-platform synchronization.

\section{Related Work} \label{sec:relatedwork}
\subsection{Challenges of Professional Software}
Professional software has long been recognized as requiring a steep learning curve. A primary challenge is the complexity of the interfaces presenting numerous icons, tool palettes, and menu commands~\cite{chilana2018supporting}. Prior work has identified that beginners frequently have trouble finding functions, understanding specialized terms, and determining the correct operational sequence within the interface~\cite{novick2009micro, lee2010usability, kiani2019beyond}. 
Another challenge is that user tasks are often open-ended and creative, and the software provides limited guidance on how to proceed. Beginners are known to feel lost about which tools to use for their goal. However, the learning resources for professional software are mainly external, e.g., formal manuals~\cite{novick2006don}, tutorial videos~\cite{lafreniere2013community}, blogs, and dedicated forums, which may disrupt workflows and make it challenging to locate information relevant to a user's immediate problems, creating a high barrier to self-taught exploration~\cite{rieman1996field, chilana2012lemonaid, chilana2018supporting, khurana2024and}. 
In summary, the key challenges of professional software include \textbf{C1: cognitive overload from extensive features at once}; \textbf{C2: limited contextual, step-by-step built-in guidance}; \textbf{C3: unfamiliar domain terminology}. 

\subsection{Simplified Alternatives for Professional Software}
To tackle these challenges, a common approach is to offer separate modes or editions for beginners and experts, which addresses \textbf{C1}. For example, beginner-friendly applications with simplified functionality have been designed to target newcomers for basic operations, e.g., Adobe Express~\cite{adobeexpress}, Canva~\cite{canva}, OpenShot~\cite{openshot}, TinkerCAD~\cite{tinkercad}. In addition, recent work has explored natural language interfaces~\cite{generativefill, qin2024instructvid2vid, poole2022dreamfusion, liu2024dancegen} or LLM-generated interfaces~\cite{vaithilingam2024dynavis, liu2025crowdgenui}, with simplified interaction modality like text or on-demand controls and widgets. 
These simplified alternatives embody the classic simplicity-power tradeoff~\cite{uiandvis1999hearst}. Simplified interfaces are easier to learn and less intimidating for novices, but they often lack the breadth of functionality that experts require. To balance this tradeoff, prior research has explored the transition between multiple interface modes, such as multi-layer interfaces~\cite{shneiderman2002promoting, mcgrenere2002evaluation} and workflow documentation interfaces~\cite{grossman2010chronicle}. 

Nonetheless, the gap between simplified and advanced interfaces may lead to new challenges: users can get stuck in the beginner mode if the two modes are too different~\cite{cockburn2014supporting}, or get overwhelmed if advancing to complex modes is poorly managed~\cite{nicol2024psychology}. Another issue is that context switching from novice to professional requires users to relearn workflows because the interfaces differ. Such discontinuity can void some of the initial learning done in the simplified application~\cite{li2022design}. 
Our work aims to solve these limitations by embedding a scaffolded interface into professional software in order to reinforce connection with the native software and reduce discontinuity from platform switching. 
To help users move from basic to advanced, we offer user-selectable complexity levels that let them gradually take on more advanced tools as they learn and work through each task.

\subsection{In-Context Assistance for Professional Software}
To address \textbf{C2}, research and development efforts have created in-context assistance in professional software. Unlike external resources, such assistance aims to provide support at the user's point of need, thus minimizing disruption and leveraging the immediate task situation. For instance, early work designed elements like tooltips to provide brief descriptions or help buttons linked to relevant sections of a manual~\cite{farkas1993role, kelleher2005stencils, grossman2010toolclips, fourney2014intertwine, photoshop}. While integrated into the application, these elements are static and often lack true context sensitivity. 

Recognizing the limitations of static help, researchers have explored more interactive and context-aware approaches. One thread of work explored \textit{intelligent user assistance}, such as using Bayesian probability networks to model the inherent uncertainty in understanding user goals and offer help without an explicit user request~\cite{li2005active, horvitz2013lumiere}. Another line of work focused on improving \textit{contextual assistance delivery} by bringing external resources into the application context, such as automatically identifying and presenting relevant web pages, documentation sections, forums, or tutorial videos based on the user's current activity~\cite{chilana2012lemonaid, fraser2019replay, matejka2011ambient, matejka2011ip, ekstrand2011searching, zhong2021helpviz, yang2022softvideo, yang2024aqua}. 

However, existing in-context assistance faces challenges in accurately inferring user goals and risks of providing irrelevant or poorly timed assistance~\cite{brdnik2022intelligent}. Second, while contextual assistance delivery reduces the need for explicit searching, users can still struggle to map the provided guidance (textual steps or visual cues) onto the complex interface of professional software~\cite{kosch2023survey}. Lastly, even with contextually retrieved resources, users may still face information overload or difficulty identifying the most relevant information~\cite{lim2012improving}.
We tackle these challenges by building a task-aware, scaffolded UI around a user's task. We select and surface only the relevant tools to avoid presenting contextually irrelevant support. Rather than making users interpret instructions and navigate a complex UI, our scaffolded interface embeds the guidance by grouping tools into workflow stages, ensuring clear task-tool alignment.

\begin{figure*}[!t]
    \centering
    \includegraphics[width=\textwidth]{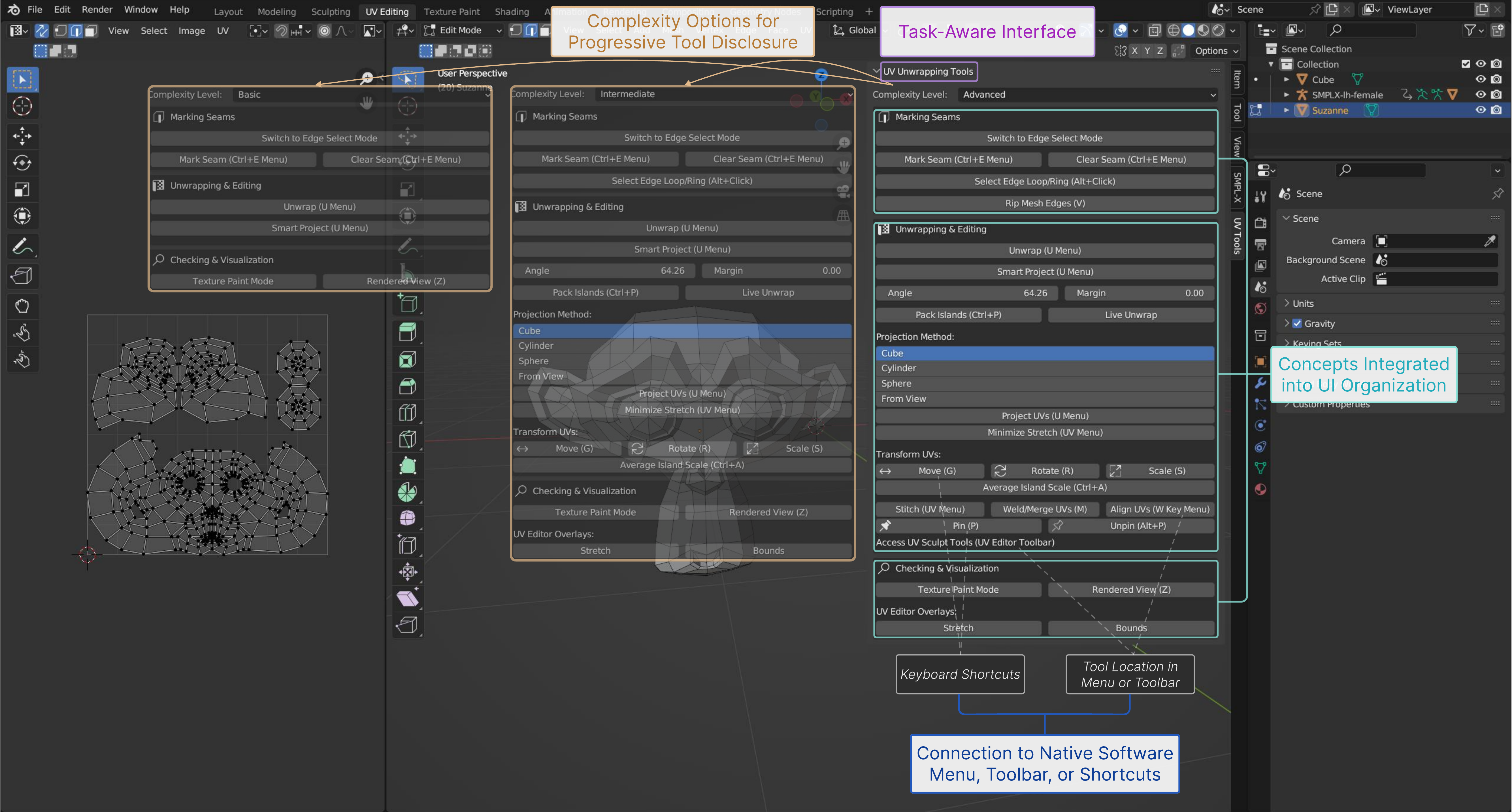}
    \caption{\papername{} implemented as a custom panel in Blender for the UV unwrapping task. Task-relevant tools are grouped into logical sections that align with workflow stages and incorporate domain concepts. Each tool includes a descriptive label and a tooltip explaining its function and how it relates to domain concepts. To help users connect the scaffolded interface with Blender’s native UI, the interface displays keyboard shortcuts and hints about where each tool can be found in Blender’s standard menus or toolbars. The interface manages complexity via user-selectable levels for progressive tool disclosure.}
    \label{fig:task1_interface}
\end{figure*}

\subsection{In-Application Software Copilots}
The recent advancement of LLMs has reshaped the landscape of in-context assistance, bringing the emergence of in-application copilots. These copilots can process natural language queries, generate human-like responses, and automate software tasks. Examples like Adobe Firefly~\cite{firefly} and Figma AI~\cite{figma} represent this new generation of assistance, aiming to offer flexible and dynamically integrated support. In addition, multimodal information, such as screen monitoring, has been leveraged to enhance user contextual understanding. For instance, Microsoft Copilot Vision~\cite{krol2025copilotvision, cunningham2025copilotvision} and Google AI Studio~\cite{google2025aistudio} aim to provide guidance based on not only textual queries but also real-time visual context.

Although these copilots are helpful to automate well-defined or simple tasks, fully automated approaches can diminish the user's sense of control and ownership over the task~\cite{endsley2017here, heer2019agency, li2024user, sellen2024rise, khurana2025me}. By abstracting away the underlying steps, automation prevents users from understanding core concepts and workflows, hindering learnability (\textbf{C3}). This goes contrary to the user-preferred method of \textit{learning by doing} as studied in prior research~\cite{carroll1984training, rieman1996field, novick2009micro, kiani2019beyond}.
We address these limitations by focusing on user engagement and embedding learning directly into the workflow. Task-relevant tools are organized around domain concepts, with clear descriptions to help users understand and connect tools to those concepts.

\section{\papername{} Design and Implementation} \label{sec:system}

\begin{figure*}[!t]
    \centering
    \includegraphics[width=\textwidth]{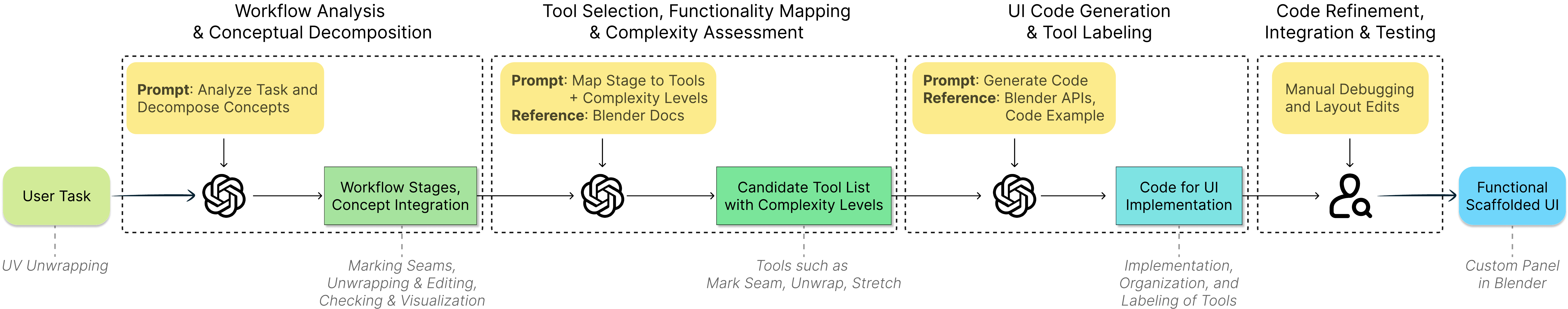}
    \caption{\papername{} technical implementation pipeline. (1) A user's task informs LLM-assisted workflow analysis to produce workflow stages that incorporate domain concepts. (2) These stages guide LLM-assisted tool selection, functional mapping, and complexity assessment, resulting in a candidate tool list. (3) The tool list is fed into LLM-based code generation, producing Python code and tool labeling for the UI implementation. (4) The UI code undergoes manual refinement, debugging, and integration as an add-on panel in Blender.}
    \label{fig:implementation}
\end{figure*}

\subsection{Design Goals and Rationale}
{\setlength{\fboxsep}{2pt}\colorbox{lightpurple}{\textbf{DG1}}}\textbf{: Task-aware interface design to reduce complexity and cognitive load}. This goal minimizes overload from excessive interface elements (\textbf{C1}) by adapting the scaffolded interface to the user’s task. The interface selects and surfaces relevant tools hidden in menus or shortcuts, which aims to lower the learning curve, reduce the need for prior knowledge, and address hidden affordances~\cite{andrade2009macro}. This is grounded in task-centric~\cite{lafreniere2014task} and adaptive design principles~\cite{knowledge_navigator, jameson2007adaptive} to manage interface complexity and reduce cognitive load~\cite{sweller2011cognitive}. 

{\setlength{\fboxsep}{2pt}\colorbox{lightorange}{\textbf{DG2}}}\textbf{: Progressive tool disclosure to scaffold learning}. This goal addresses the lack of step-by-step in-context guidance (\textbf{C2}) by applying instructional scaffolding~\cite{brusilovsky2007user, belland2017instructional}. Task-relevant tools are revealed by complexity level: basic tools appear first, with advanced tools disclosed gradually through user-selectable complexity levels. This aligns with the training-wheels interface model~\cite{carroll1984training} and intermodal guidelines~\cite{cockburn2014supporting} to manage learning and encourage exploration.

{\setlength{\fboxsep}{2pt}\colorbox{lightcyan}{\textbf{DG3}}}\textbf{: Integrating domain concepts into interface design for workflow clarity and learning}. This goal tackles unfamiliar terminology (\textbf{C3}) by organizing tools according to task workflow stages and using descriptive labels to link tools to domain concepts. It connects software operations, domain terminology, and task workflow, which aims to support conceptual learning during use. This is based on constructivist learning theory~\cite{papert2020mindstorms}, research on learning by doing~\cite{rieman1996field, novick2009micro}, and vocabulary extension~\cite{cockburn2014supporting}.

{\setlength{\fboxsep}{2pt}\colorbox{lightblue}{\textbf{DG4}}}\textbf{: Linking the scaffolded interface to native software for learning transfer}. This goal addresses potential overreliance on the scaffolded interface by connecting the selected tools back to their native counterparts. Motivated by distributed cognition theory~\cite{hollan2000distributed}, the scaffolded interface uses tooltips and labels to show associated shortcuts and menu paths, helping users connect to the native software and build long-term software fluency.

\subsection{Usage Scenario}
UV unwrapping is a fundamental process for applying 2D textures to a 3D model's surface. It involves \textit{cutting} and \textit{unfolding} the 3D mesh into a flat 2D representation called a UV map. Each point on this 2D map (defined by U and V coordinates) corresponds to a vertex on the 3D model, allowing textures to be projected onto its surface. The basic process includes defining seams -- edges on the 3D model where the mesh will be split, then executing an unwrap operation to generate the 2D layout of UV islands. Thereafter, these islands are arranged and optimized in the 2D space, and finally, the layout is checked for distortions or overlaps. 

Performing this task in Blender requires navigating different modes and menus. Users first enter Edit Mode, manually select edges, and mark them as seams via context menus like Ctrl+E or dedicated UV menus. After this step, an unwrapping operator from the U menu is applied. The resulting UVs are then viewed and edited in a separate UV Editor workspace. Users need to locate and utilize various tools accessed through keyboard shortcuts, specific UV menu options, or the UV Editor's toolbar to arrange, scale, and optimize the UV layout. 

Using our scaffolded interface (Figure~\ref{fig:task1_interface}) to work on this task, a user starts at the \textsc{Basic} complexity level and sees a tailored workflow integrated with domain concepts. They first work through the \textit{Marking Seams} stage using fundamental tools. Following the suggested flow, they move to \textit{Unwrapping \& Editing} to unwrap the UV map and perform a simple visual check using tools in \textit{Checking \& Visualization}. Seeking to improve this initial result, they switch the interface's complexity level to \textsc{Intermediate} to utilize newly available seam selection aids to refine the UV layout. For final polishing and access to the complete toolset relevant to this task, they select the \textsc{Advanced} level to reveal specialized tools and employ advanced operators for accurate adjustments, and analysis features to minimize distortion and verify the UV map quality. Domain concepts (e.g., seams, UV islands, stretch) are clearly labeled with further details available via mouseover tooltips.

\subsection{Implementation Details}
Building upon the design goals, we propose a technical pipeline and implement the scaffolded interface in Blender (version 3.6) using its Python API for scripting. The implementation involves using an LLM (GPT-4o) to assist with task analysis, tool selection, and code generation. The complete prompt is provided in Appendix \ref{sec:llm_prompt}, and the implementation pipeline is shown in Figure~\ref{fig:implementation}. 

\textbf{Workflow analysis and conceptual decomposition}: The pipeline begins with workflow analysis and conceptual decomposition facilitated by LLM reasoning. A user's task is provided in input prompts. The LLM helps identify and organize sub-tasks that define the structure and navigation of the interface. At the same time, the LLM breaks down domain concepts linked to each stage. These concepts are presented to users through textual cues and tooltips, guiding task execution and supporting concept understanding.

\textbf{Tool selection, functionality mapping, and complexity assessment}: Subsequently, the LLM is used to select tools and map them to relevant functionalities. For each workflow stage, prompts are formulated to query the LLM for relevant Blender Python API functionalities, including operators, properties, and existing UI elements tied to specific actions. As part of this process, the LLM also suggests a complexity level for each tool (e.g., \textsc{Basic}, \textsc{Intermediate}, \textsc{Advanced}), based on typical Blender usage patterns and related data. These complexity levels are then used to implement progressive tool disclosure in the interface.

\textbf{UI code generation and tool labeling}: Next, Python scripts for the interface implementation are generated using LLM-based code generation. These include the main UI container classes and associated classes for storing settings, such as tool complexity levels. Prompts provided to the LLM include the workflow stages, conceptual decomposition, the list of tools with their assigned complexity levels, the desired interface layout hierarchy, and the logic to support progressive disclosure based on complexity.

\textbf{Code refinement, integration, and testing}: The LLM-generated Python code is manually reviewed, debugged, and executed in Blender to create a custom add-on panel. The visual organization and functional behavior of all UI elements are manually tested in Blender to ensure correctness and usability. 

\subsubsection{Advantages of LLM-Assisted Implementation}
Using an LLM helped to accelerate our implementation by leveraging broad training data to rapidly generate reasonable workflows and conceptual decompositions. The LLM also suggested API functions, aiding the discovery of relevant APIs offered by Blender. For interface implementation, LLM-based code generation helped generate UI code from structured specifications. During this implementation process, while human involvement remains crucial for validation, prompt refinement, and API verification, our pipeline enabled fast prototyping and iteration cycles and might be valuable for researchers and developers who aim to develop new scaffolded interfaces.

\subsubsection{Broader Applicability and Customization}
Although we introduce a specific implementation pipeline, the underlying methodology has broader applicability and customizability. The principles of task decomposition, concept-driven UI organization, and progressive disclosure can be applied to other software platforms with programming APIs for UI and functionality implementation (e.g., Maya~\cite{maya_api}, AutoCAD~\cite{autocad_api}, Unity~\cite{unity_api}). The pipeline components, such as the workflows, tool selections, and complexity levels, are highly customizable by designers. Therefore, this implementation serves as a concrete realization of our interface design method, suitable for supporting the creation of scaffolded interfaces for a variety of tasks, platforms, and user needs (See further details in Sections~\ref{sec:results_generalizability} and~\ref{sec:discussion_generalizability}).

\section{User Study Design}

To evaluate the implemented scaffolded interface, we conducted user studies with beginners (Study 1) and experts (Study 2). This section introduces the detailed study design.

\subsection{Study 1: Beginner Users}
\subsubsection{Study Goals}
From the study with beginners, we aim to explore the following research questions (RQs) geared to evaluate the effectiveness of our interface compared to the baseline interface, i.e., the default Blender interface. 

\textbf{RQ1: How does the scaffolded interface influence task performance?} This question explores whether creating task-aware interfaces with relevant tools (\textbf{DG1}) impacts measurable aspects like perceived task load, workflow clarity, task performance, task completion time (efficiency), and user confidence and engagement.

\textbf{RQ2: How does the scaffolded interface affect the learning of domain concepts and software operations?} This question evaluates how progressive tool disclosure (\textbf{DG2}), concept-integrated UI organization (\textbf{DG3}), and bridging to native software (\textbf{DG4}) help users understand domain concepts, connect these concepts to specific tools, and learn the native software operations.

\textbf{RQ3: How does the user experience differ with the scaffolded interface?} This question investigates the overall user experience, including perceived ease of use, sense of support vs. constraint, entry barriers, and reliance on external help and documentation.

\subsubsection{Experimental Design}
We adopted a between-subjects design where each participant was randomly assigned to one of two interface conditions, the default Blender interface or our interface, and to one of two task conditions. Task 1 asked participants to perform UV unwrapping on a cube, a beginner-level texturing task, while Task 2 involved constructing a human walk cycle, a basic animation task. We provided all participants with the same cube for Task 1 and the same rigged human character ready for animation for Task 2 to ensure experiment consistency. The interfaces for Tasks 1 and 2 are shown in Figure~\ref{fig:task1_interface} and Appendix~\ref{sec:task2_interface}, respectively. 
To ensure consistent study instruction, we provided a study guide (Appendix~\ref{sec:study_guide}) that introduced the basics of UV unwrapping and animation, as well as publicly available tutorial videos, with durations of 4.5 minutes for Task 1 and 7 minutes for Task 2. Task 2 was more complex than Task 1 due to the longer completion time and more required steps. 

\subsubsection{Participant Recruitment}
We recruited 32 English-speaking beginners (11 female, 19 male, 2 non-binary; mean age = 27.31, SD = 7.34) with no or limited 3D modeling and animation experience via Prolific (100\% approval rate)~\cite{prolific}. Participants were randomly assigned in a fully balanced design to one of two interface conditions, (1) Baseline (default Blender), or (2) Ours (scaffolded interface), to complete either Task 1 or Task 2, resulting in eight participants per task-interface combination.

\subsubsection{Measurements} \label{sec:study1_measurements}
To evaluate participant experience, we used four questionnaires with 7-point Likert scale responses. 
For RQ1, we used the NASA Task Load Index (TLX)~\cite{hart2006nasa} and a custom task performance questionnaire. For RQ2 and RQ3, we used custom questionnaires on concept learning and user interface experience. 
These custom questionnaires were designed based on broadly used standard questionnaires and an instructional model, including Software Usability Measurement Inventory (SUMI)~\cite{kirakowski1996software}, Instructional Materials Motivation Survey (IMMS)~\cite{keller2010imms}, and ARCS model of motivation~\cite{keller2009motivational}. All these questionnaires are available in Appendix~\ref{sec:study1_questionnaires}. 

\subsubsection{Procedure}
We invited eligible participants with no Blender experience for 3D modeling or animation tasks to a Zoom meeting. Before the study started, we emailed a consent form (IRB protocol \# anonymized), which outlined the study goal, procedure, duration, and data collection. After providing informed verbal consent, participants filled out a pre-study demographics questionnaire, which took around 5 minutes.

We then shared a study guide via a Google Slides link in Zoom chat, which included task instructions and tutorials. Once participants confirmed they had read the guide and accurately described the task, we shared our desktop screen, gave them remote control, and asked them to complete their assigned task. With consent, we recorded the desktop screen to record their task actions. Participants were encouraged to think aloud, and we took notes when observing them working on their tasks. Based on a pilot study with two beginner users that helped us estimate task completion time, we gave the participants 20 minutes for Task 1 and 30 minutes for Task 2. 

After the task, participants were invited to complete the post-study questionnaires, which took around 10 minutes. The study concluded with a payment of 10 USD to participants who worked on Task 1 for completing the 30-35 minute study and 15 USD to participants who worked on Task 2 for completing the 40-45 minute study. 

\subsubsection{Data Analysis}
To evaluate the effects of interface and task on participant responses, we conducted Type II two-way analyses of variance (ANOVA)~\cite{seabold2010statsmodels} with \textit{Interface} (Ours vs. Baseline) and \textit{Task} (Task 1 vs. Task 2) as between-subjects factors to examine the main and interaction effects (\textit{Interface} $\times$ \textit{Task}) on our measurements. Following significant ANOVA effects, we performed post-hoc pairwise comparisons using Bonferroni-adjusted t-tests~\cite{Vallat2018pingouin} to identify specific differences.

\subsection{Study 2: Expert Users}
\subsubsection{Study Goals}
For Study 2, we kept \textbf{RQ1--RQ3} for a comparative analysis across beginner and expert users. In addition, two further research questions are formulated for the experts to explore aspects of skill acquisition and the generalizability of our method.

\textbf{RQ4: How can the scaffolded interface potentially shape long-term skill development?} This question explores expert perspectives on whether the scaffolded interface could help refine skills, adapt tools to new tasks, and strengthen concept understanding for solving new problems. We were interested in their qualitative insights into the impact of our scaffolded interfaces on long-term skill acquisition.

\textbf{RQ5: What is the potential for applying the scaffolded interface design method to other software?} This question asks for qualitative insights from experts familiar with additional professional software beyond Blender. We were interested in their perspectives on the feasibility, benefits, and challenges of using our scaffolded interface design method across different platforms and domains.

\subsubsection{Experimental Design}
Each participant was assigned to one of two task conditions. Task 1 involved performing UV unwrapping on a 3D monkey head (Figure~\ref{fig:task1_interface}), an advanced texturing task, while Task 2 required constructing a human walk cycle that reflected personality, an open-ended animation task. We provided the 3D model for Task 1 and the rigged human character for Task 2 for experiment consistency. 
Task assignment was based on participant prior experience: only those with 3D model texturing experience were assigned to Task 1, and only experienced animators were assigned to Task 2. This allowed us to gather feedback grounded in a participant's prior expertise to enable comparisons between their experience with the Blender interface and our interface.

\subsubsection{Participant Recruitment}
We recruited eight experts (3 female, 5 male; mean age = 28.87, SD = 3.66) via Prolific~\cite{prolific}, requiring at least one year of Blender experience and preferring additional proficiency in tools like Maya or 3ds Max. All other criteria for gender, age, language, and Prolific history matched those of Study 1. Participants (S2-P1 to S2-P8) were each assigned to either Task 1 or Task 2 and completed their task using only our scaffolded interface, with four experts per task. Detailed demographics appear in Appendix~\ref{sec:study2_demographics}.

\subsubsection{Measurements}
To evaluate RQs 1, 2, and 3, we used the same questionnaires as Study 1 but adapted the items to preference-based questions: each item was rated on a 7-point Likert scale (1 = strongly prefer our interface, 7 = strongly prefer the baseline interface). Additionally, we collected qualitative feedback by asking participants to elaborate on their experience with our interface regarding task performance, concept learning, and overall interface experience. 
For RQs 4 and 5, we gathered qualitative data in the form of participant feedback regarding skill development and the potential for generalizing the scaffolded interface approach to other software tools. Refer to Appendix~\ref{sec:study2_openended_questions} for the complete set of open-ended questions for qualitative data collection.

\subsubsection{Procedure}
Once users responded to our recruitment call, we invited eligible participants to our study via Zoom. The pre-study session involving the consent and pre-study questionnaire had the same procedure as Study 1. 
Following this, we offered an introduction to our interface, during which we shared our desktop screen with the scaffolded interface shown in Blender and introduced how the interface was designed for their task, covering all the design goals. This step took around 5 minutes. 
Next, participants used our interface to work on either the UV unwrapping or animation task for 15 minutes through remote desktop control. We recorded the desktop screen with participant consent and took notes on their actions. 
After the task, participants completed the post-study questionnaires and answered the open-ended questions, which took around 10 minutes. The study concluded with a payment of 10 USD for completing the 30-minute study. 

\subsubsection{Data Analysis}
The qualitative data was analyzed using thematic analysis~\cite{braun2006using, braun2012thematic}. The authors reviewed the participant responses, coded a subset of the data to extract relevant information, refined the codes, and categorized themes centered around the research questions.

\begin{figure*}[!ht]
    \centering
    \subfloat[Task Load]{
        \centering
        \includegraphics[width=0.486\textwidth]{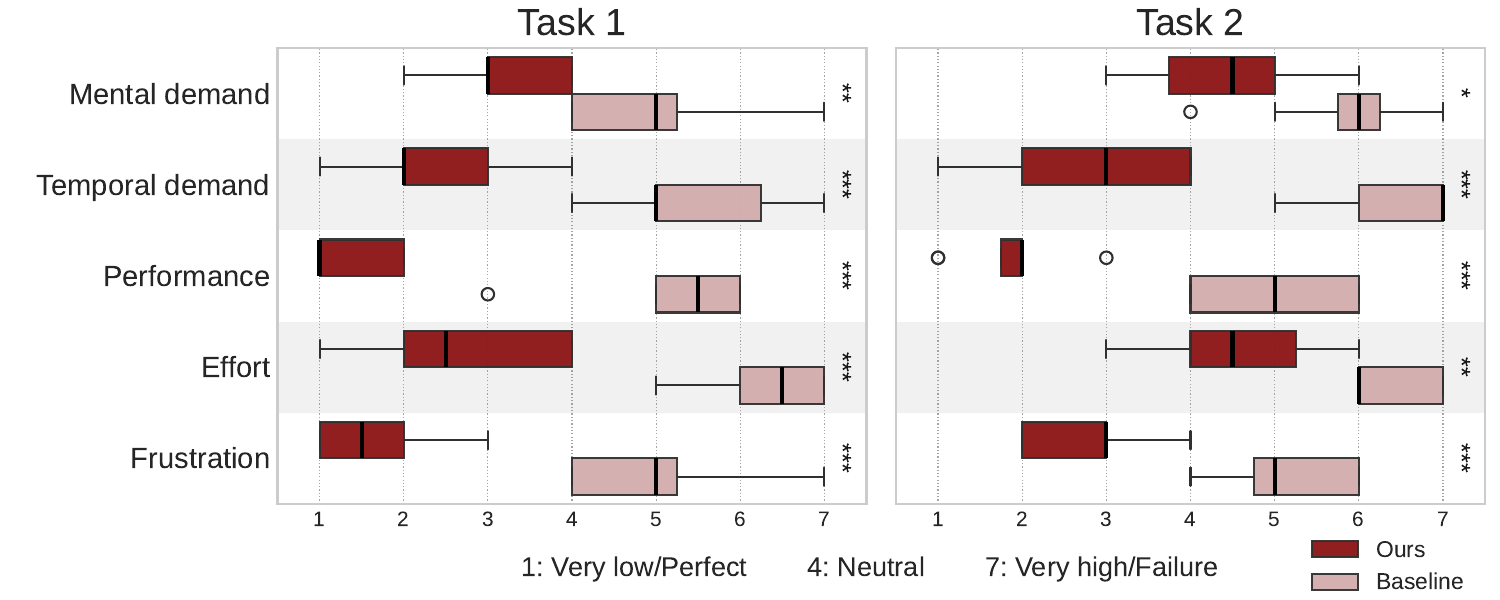}
        \label{fig:study1_cognitive_load}
    }
    \hfill
    \subfloat[Task Performance]{
        \centering
        \includegraphics[width=0.486\textwidth]{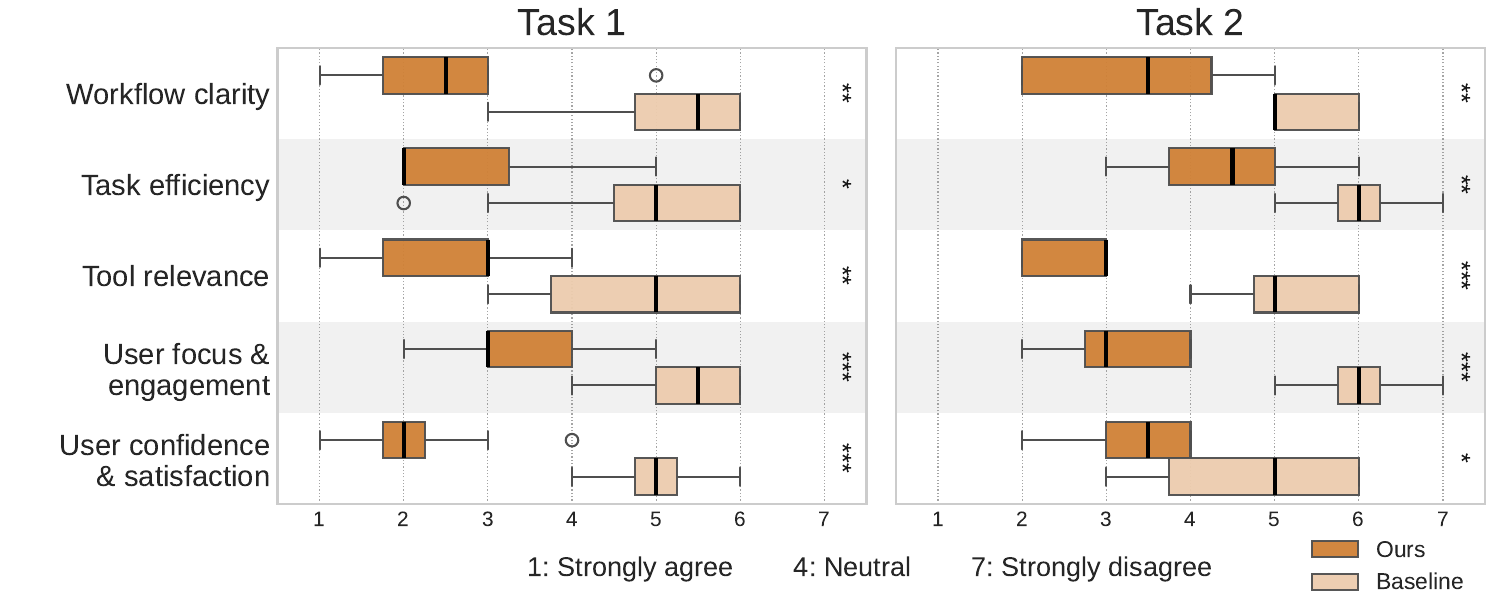}
        \label{fig:study1_task_performance}
    }
    \hfill
    \subfloat[Concept Learning]{
        \centering
        \includegraphics[width=0.486\textwidth]{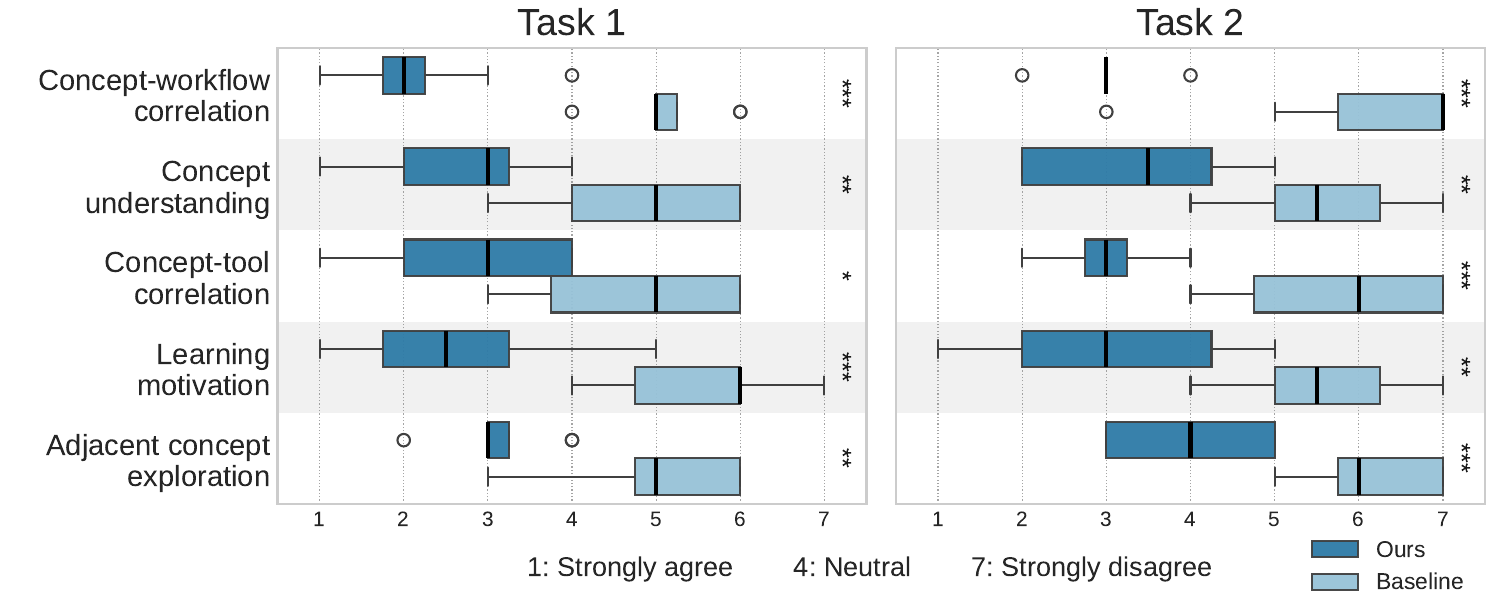}
        \label{fig:study1_concept_learning}
    }
    \hfill
    \subfloat[User Interface Experience]{
        \centering
        \includegraphics[width=0.486\textwidth]{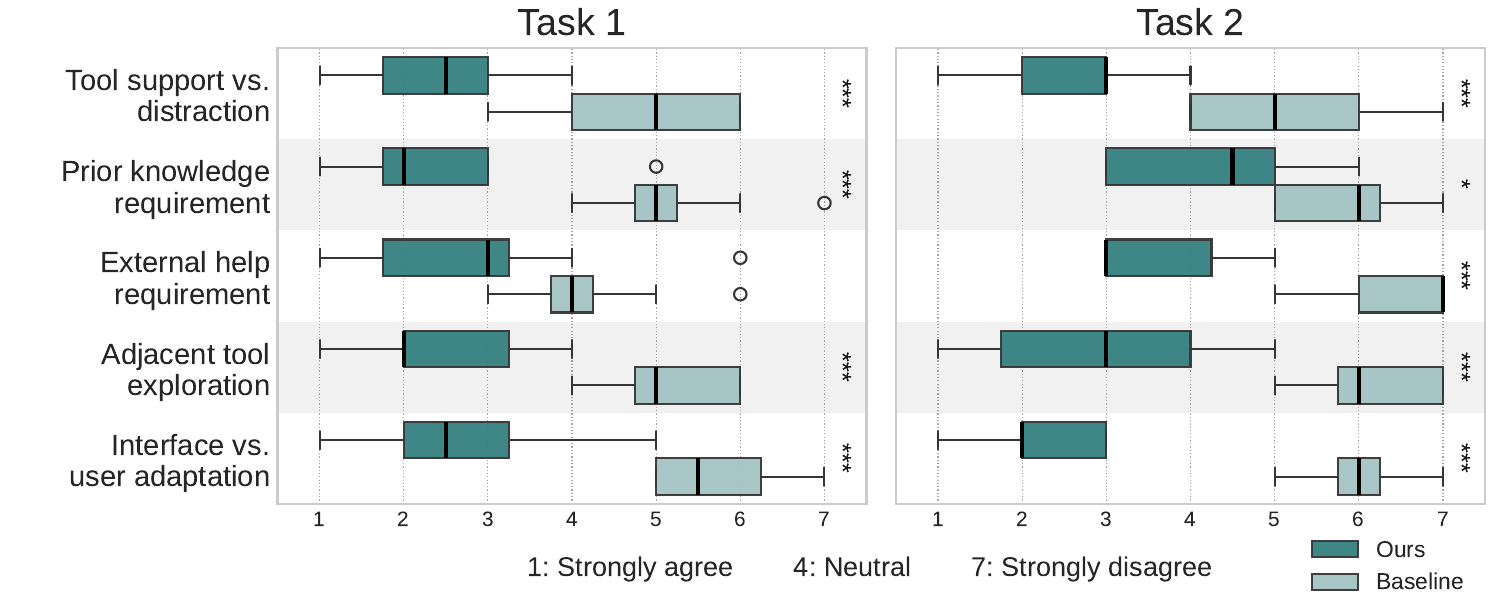}
        \label{fig:study1_interface_experience}
    }
    \caption{Study 1 participant responses using Ours and Baseline interfaces for Tasks 1 and 2 (lower is better). Asterisks denote significant differences in post-hoc pairwise comparisons between interfaces within each measure ($*$: $p < .05$, $**$: $p < .01$, and $***$: $p < .001$).}
    \label{fig:study1_measurements}
\end{figure*}

\section{User Study Results}

\subsection{Study 1: Beginner Users}
\subsubsection{Questionnaire Responses}
Figure~\ref{fig:study1_measurements} shows the participant self-reported responses. Asterisks in the figure indicate the statistical significance derived from post-hoc pairwise comparisons between Ours and Baseline interfaces for each task and measure. The complete two-way ANOVA results are reported in Appendix~\ref{sec:study1_anova}.

\paragraph{Task Load}
A significant main effect of \textit{Interface} was observed on all five dimensions: mental demand ($F_{(1, 30)}=22.53, p<.001$), temporal demand ($F_{(1, 30)}=95.38, p<.001$), performance ($F_{(1, 30)}=150.36, p<.001$), effort ($F_{(1, 30)}=69.96, p<.001$), and frustration ($F_{(1, 30)}=91.43, p<.001$). 
A significant main effect of \textit{Task} was found for mental demand ($F_{(1, 30)}=8.53, p<.01$), temporal demand ($F_{(1, 30)}=4.71, p<.05$), effort ($F_{(1, 30)}=8.51, p<.01$), and frustration ($F_{(1, 30)}=4.32, p<.05$), indicating higher perceived load on these dimensions in Task 2. 
A significant interaction effect was found for effort ($F_{(1, 30)}=8.51, p<.01$). 
The post-hoc pairwise comparisons, as shown in Figure~\ref{fig:study1_cognitive_load}, revealed that participants using our interface reported significantly lower task load for both tasks. 
These results respond to DG1 and RQ1, confirming that our interface helps decrease user-perceived task load by reducing interface complexity. 

\paragraph{Task Performance}
A significant main effect of \textit{Interface} was seen on all five measures: workflow clarity ($F_{(1, 30)}=34.60, p<.001$), task efficiency ($F_{(1, 30)}=19.96, p<.001$), tool relevance ($F_{(1, 30)}=48.13, p<.001$), focus and engagement ($F_{(1, 30)}=71.46, p<.001$), and confidence and satisfaction ($F_{(1, 30)}=38.53, p<.001$). 
A significant main effect of \textit{Task} was found for task efficiency ($F_{(1, 30)}=12.55, p<.001$). 
A significant interaction effect was observed for confidence and satisfaction ($F_{(1, 30)}=4.80, p<.05$).
Post-hoc comparisons, as reported in Figure~\ref{fig:study1_task_performance}, indicate that our interface has significantly outperformed the baseline on all five measures in both tasks. These results respond to DG1 and RQ1, revealing our interface's positive impact on task completion assistance. 

\paragraph{Concept Learning}
A significant main effect of \textit{Interface} was shown across all measures: concept-workflow correlation ($F_{(1, 30)}=78.91, p<.001$), concept understanding ($F_{(1, 30)}=29.67, p<.001$), concept-tool correlation ($F_{(1, 30)}=33.39, p<.001$), learning motivation ($F_{(1, 30)}=36.25, p<.001$), and adjacent concept exploration ($F_{(1, 30)}=41.20, p<.001$). 
A significant main effect of \textit{Task} was observed for concept-workflow correlation ($F_{(1, 30)}=7.39, p<.05$) and adjacent concept exploration ($F_{(1, 30)}=10.30, p<.01$). No significant interaction effect emerged.
Post-hoc comparisons, as seen in Figure~\ref{fig:study1_concept_learning}, showed that our interface significantly improved concept and tool learning across all measures. These results, in response to DG2--4 and RQ2, show that our interface helps users learn task-related concepts, explore adjacent concepts, and stay motivated to learn. 

\paragraph{User Interface Experience}
A significant main effect of \textit{Interface} was observed across all dimensions: tool distraction vs. support ($F_{(1, 30)}=44.44, p<.001$), prior knowledge requirement ($F_{(1, 30)}=32.36, p<.001$), external help requirement ($F_{(1, 30)}=26.75, p<.001$), adjacent tool exploration ($F_{(1, 30)}=59.25, p<.01$), and user vs. interface adaptation ($F_{(1, 30)}=104.14, p<.001$). 
A significant main effect of \textit{Task} was found for prior knowledge requirement ($F_{(1, 30)}=11.65, p<.01$) and external help requirement ($F_{(1, 30)}=15.35, p<.001$). No significant interaction effect was observed. 
Post-hoc comparisons, as displayed in Figure~\ref{fig:study1_interface_experience}, showed that our interface significantly improved interface experience across all five dimensions in both tasks, except for the external help requirement in Task 1. These results, in response to RQ3, suggest that our interface augments user experience by reducing their reliance on prior knowledge or external guidance, lowering distraction, and supporting tool exploration.

\begin{figure*}[!ht]
    \centering
    \includegraphics[width=0.75\textwidth]{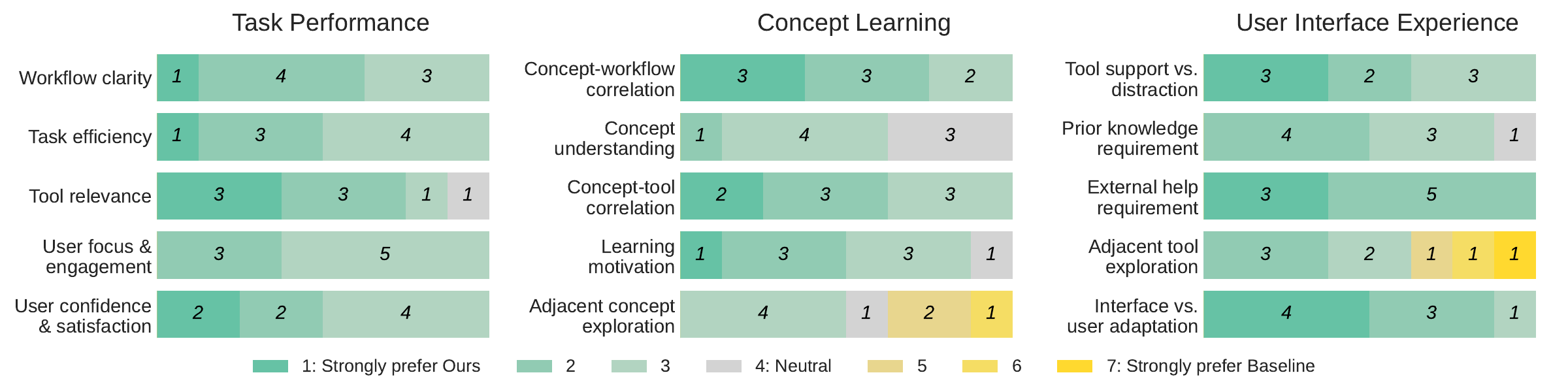}
    \caption{Study 2 participant responses for task performance, concept learning, and user interface experience comparing Ours and Baseline interfaces.}
    \label{fig:study2_scores}
\end{figure*}

\subsubsection{Task Completion Time}
All participants successfully completed their tasks and generally spent less time using our interface compared to the baseline: 
\begin{align*}
    \bar{t}_{Ours (Task 1)} &= 12.69 \pm 1.49 \textrm{ mins}, \\ \bar{t}_{Baseline (Task 1)} &= 15.08 \pm 2.34 \textrm{ mins}; \\
    \bar{t}_{Ours (Task 2)} &= 20.89 \pm 2.09 \textrm{ mins}, \\ \bar{t}_{Baseline (Task 2)} &= 25.33 \pm 2.88 \textrm{ mins}.
\end{align*}

Significant main effects of \textit{Interface} ($F_{(1, 30)}=14.78, p<.001$) and \textit{Task} ($F_{(1, 30)}=94.71, p<.001$) were observed. The interaction effect was not significant. Post-hoc analysis revealed no significant difference in completion time in Task 1, but users with our interface were significantly faster in Task 2 ($p<.05$), demonstrating its efficiency advantage primarily under higher task demands.

\subsubsection{User Behavior}
We observed that participants using our interface progressively reduced their reliance on tutorial videos and increasingly navigated via the interface's inherent workflow structure, echoed by participant self-reported improved workflow clarity, tool relevance, engagement, and reduced external help requirement. In contrast, control group participants consistently referred to the videos throughout the tasks. Consequently, the experimental group generally completed tasks faster due to reduced time spent searching for required tools. 

When certain tools were not mentioned in the tutorials but shown in our interface, participants tended to explore them during or after working on their tasks, as echoed by participant self-reported tool and concept exploration. Conversely, control group participants rarely explored beyond the tools in tutorials.

\subsection{Study 2: Expert Users}
\subsubsection{Questionnaire Responses}
Figure~\ref{fig:study2_scores} shows the participant self-reported scores for task performance, concept learning, and user interface experience measures. 
Participants had a clear preference for our interface across all dimensions regarding task performance. 
For concept learning, participants generally perceived our interface as more supportive for understanding concepts, relating tools to concepts and workflows, and motivating learning. However, they had mixed opinions on adjacent concept exploration, where 4 favored our interface, 1 was neutral, and 3 favored the baseline.
For the overall interface experience, consistent preference for our interface was observed except for adjacent tool exploration, where 5 participants preferred our interface, and 3 favored the baseline.

\subsubsection{Qualitative Results}
\paragraph{Tradeoffs of Task-Aware Interface on Task Performance and Workflow Flexibility (RQ1)}
Participants generally appreciated how our interface supported task performance by surfacing task-relevant tools. As S2-P1 noted, ``\textit{I liked how I didn’t have to dig around for modifiers. They just showed up when relevant}.'' This interface design led to higher task efficiency and user focus as S2-P2 said that, ``\textit{I actually finished the UV task faster than expected},'' S2-P7 described the experience as ``\textit{fast and easy to do basic modeling and modifiers},'' and S2-P8 echoed that ``\textit{just tools that made sense at the moment}.'' These responses suggest that organizing tools into task-aware UI has helped reduce interface distraction, support task efficiency, and augment user engagement.

However, participants also expressed concerns that the system's assumptions sometimes misaligned with their preferred workflows. Some found the interface limiting when it failed to surface expected tools at the necessary moment, e.g., S2-P4 remarked that, ``\textit{the interface seemed to over-assume what I want}.'' Others perceived the guided structure forced a specific linear process, as S2-P6 pointed out: ``\textit{the steps feel rigid that force a linear way of working}.'' This perceived rigidity consequently constrained workflow flexibility, with them saying that, ``\textit{I sometimes feel boxed in and want to change its logic}'' (S2-P6). These findings show the need for more user customization or smarter adaptive features to support expert users who prefer dynamic or non-linear workflows.

\paragraph{Concept-Integrated UI Organization Benefits Learning and Exploration but May Obscure Full Context Conprehension (RQ2)}
Participants liked our interface for its ability to clarify the relationship between concepts and tasks, which can benefit educational or onboarding contexts. For example, S2-P1 noted that, ``\textit{it'd be great for teaching junior artists. The layout helps connect tools with goals}'' and S2-P3 emphasized that, ``\textit{it makes the connection between steps and tools clear}.'' They also highlighted the learning benefits, saying that, ``\textit{it’s not just clicking through menus but also teaching you why they matter}'' (S2-P2). These results indicate the interface can support learning by linking concepts to task execution. 

In addition, progressive tool disclosure was appreciated by experts as a way to guide structured concept learning. S2-P3 noted that, ``\textit{it shows graph editing after you’re done with key posing, and this can help avoid overwhelming learners}'' and S2-P7 commented that, ``\textit{it didn’t feel like a tutorial, just well-paced exposure to new tools},'' supporting the idea that the interface encourages learning without being intrusive.
Participants also remarked that showing tools adjacent to the immediate tasks was helpful to motivate learning. As noted by S2-P5, ``\textit{It prompted me to notice things like secondary motion I’d normally skip past},'' indicating that the interface enables the discovery of related concepts and tools and encourages users to deepen their exploration beyond the task at hand. 

Despite these benefits, some participants found that the guided structure might limit system comprehension. S2-P1 commented, ``\textit{I missed seeing the full context sometimes, like how different parts of Blender connect}.'' S2-P6 similarly observed, ``\textit{you lose the bigger picture of how Blender’s laid out},'' expressing concern that the curated experience might hinder holistic interface understanding. Moreover, S2-P7 added that the interface ``\textit{didn’t explain why some tools were shown and not others}.'' Although the tools were shown and organized based on the corresponding task, this comment revealed a potential transparency gap in the scaffolded interface’s logic. 

\paragraph{Gains of Simplicity, Clarity and Focus, and Challenges of Familiarity in User Interface Experience (RQ3)}
Participants responded positively to the overall interface experience. For example, S2-P1 praised it as ``\textit{clean, not overwhelming},'' S2-P4 described the interface as ``\textit{task-centered},'' S2-P7 commented that the interface ``\textit{felt lightweight and approachable},'' and S2-P8 mentioned it ``\textit{limited distractions}.'' These results reflect participant appreciation for the interface's task-focused, clear, and simple design. S2-P2 also noted that, ``\textit{I can imagine a less fighting the UI if I were a newbie},'' indicating reduced cognitive friction during tool navigation for beginners. 

Nevertheless, participants commented on how the scaffolded interface sometimes conflicted with their familiarity with the standard software layout. S2-P3 remarked that they ``\textit{missed the old familiarity},'' pointing to the inherent tension when introducing novel interfaces. This was echoed by S2-P4, who found it ``\textit{slightly frustrating when I want to jump around between workflows quickly},'' suggesting that the structured guidance could hinder established working habits of experts. 

\paragraph{Scaffolding can Play an Effective Role for Skill Development (RQ4)}
Participants generally found that our interface provided a supportive foundation for building practical skills, particularly for users newer to Blender. S2-P1 commented, ``\textit{over time, that builds a mental model you can carry with you},'' S2-P5 called it a ``\textit{great stepping stone from beginner to intermediate},'' and S2-P6 noted the interface made it ``\textit{easier to build confidence in doing simple stuff}.'' Together, these responses show that experts see our interface as helping users build confidence, develop mental models, and gain skills gradually while using professional software.

\paragraph{Generalizability of Scaffolded Interfaces based on Transferable Concepts (RQ5)} \label{sec:results_generalizability}
Participants with experience in other professional software recognized the underlying principles of scaffolded interface design as being transferable. For instance, S2-P1 commented that, ``\textit{the tool groups it uses are transferable for folks moving between platforms}'' and S2-P8 stated that, ``\textit{I think the knowledge about motion, timing, bone constraints translates to other tools}.'' 

Building on this perceived transferability, participants suggested potential applications and impacts of generalizing the scaffolded interface design to other software. S2-P6 said that, ``\textit{3ds Max can benefit from this. The modifier stack alone is complex enough. If tools were grouped by task and revealed based on what you're actually doing, it’d cut down on confusion}.'' Others proposed similar context-aware adaptations, suggesting that, ``\textit{systems only show rigging tools once you add a skeleton, or animation curves once you key a pose, could make the whole thing become easier}'' (S2-P8). Participants also anticipated benefits, reflecting that, ``\textit{it could help new users get productive faster, and even experienced users might find it useful to stay focused}'' (S2-P6). Collectively, these results suggest user enthusiasm for applying our interface design method more broadly to reduce interface complexity and enhance workflow efficiency.

\subsubsection{User Behavior}
We observed that expert users approached tasks flexibly, sometimes bypassing the step-by-step workflow stages in our interface. For instance, S2-P6 switched between marking seams (stage 1) and unwrapping \& editing (stage 2). They started with the ``Select Edge Loops'' tool in stage 1 to mark seams, used ``Unwrap'' in stage 2 to check the UV map, then went back to use ``Edge Select'' in stage 1 to refine the seams and repeated the process. This reflects their preference for multitasking and non-linear workflows. 

When needed tools were not presented at their current complexity level, several participants (S2-P2, S2-P5, S2-P6, S2-P8) sometimes did not look for these tools by advancing complexity levels. Instead, they accessed tools via keyboard shortcuts or the default Blender interface -- the operations they were familiar with. This aligns with their feedback that the scaffolded interface was less supportive for tool exploration compared to the experience of beginners.

\section{Discussion}
\subsection{Interfaces with Task-Sensitive Adaptation and Granular Customization to Support Learning and Teaching}
Beginner participants reported benefits in both task performance and concept learning with our scaffolded interface, showing its promise for instructional use. To better support learning, the interface should adapt its scaffolding based on task complexity. 
For simpler tasks, the interface can serve as a guided playground. Study 1 showed that beginners felt encouraged to explore adjacent tools and concepts beyond those directly related to their task at hand. Future designs could expand the available tools slightly beyond the current step to promote curiosity-driven exploration. 
For complex tasks, the focus shifts to deep understanding and accuracy. In these cases, the interface can offer a smaller, essential toolset with clearer links between domain concepts and software operations. It could also include best-practice tips or visual aids like video clips. This tighter guidance can help ensure key concepts and operations are grasped before moving on. 

In addition, empowering educators is essential to unlock the full potential of scaffolded interfaces for instruction. Instructors could choose which concepts and tools the interface highlights, tailoring it to their teaching goals. This can let them design learning modules that match their instructional approach. 
Equally important is control over the learning process. Expert participants noted that progressive tool disclosure helps reduce overwhelm and foster skill development. If instructors set when new tools and concepts appear, they can pace learning based on student progress, creating a structured path that builds complexity gradually. 
By offering this blend of task-sensitive adaptation for learners and granular customization for instructors, scaffolded interfaces can evolve into helpful assistance in the educational process. 

\subsection{Scaffolded Interfaces Driven by Productivity and Creativity}
While beginners profited from structured guidance, the needs of expert participants differed, demanding a separate perspective for scaffolded interface design. Study 2 findings revealed that experts focused on boosting productivity through personalized workflows and increasing creativity through flexible exploration. Therefore, scaffolded interfaces targeting experts should evolve beyond structured guidance towards systems driven by these needs.

A scaffolded interface for productivity could learn a user’s unique shortcuts, tool sequences, and niche commands, and then optimize workflows by bundling actions into macros, rearranging palettes for top tools, or suggesting more efficient alternatives. Implementing this requires robust user activity logging, potentially through screen monitoring used by in-application copilots~\cite{cunningham2025copilotvision, google2025aistudio} to adapt interfaces according to the user's demonstrated behavior. 

Furthermore, experts often desire to solve problems creatively, stepping outside routine workflows. An expert-focused interface can support this by acting as a serendipity engine that encourages exploration. For instance, it can suggest unusual tool pairings based on deeper function relationships, highlight rarely used but relevant tools, or show alternate workflows drawn from expert behavior. 
This can be powered by LLMs, which use the user’s current state and project context to generate uncommon ideas or workflow stages. We explore this potential by implementing scaffolded interface variations adapted to creative demands for UV unwrapping and walk cycle animation, as shown in Appendix~\ref{sec:interface_variations}. 

\subsection{A Scaffold Meta-Layer for Cross-Platform Workflows} \label{sec:discussion_generalizability}
An exciting insight from Study 2 was the potential to use the scaffolded interface design method beyond Blender (RQ5). Experts highlighted the transferrable concept-driven design and expressed enthusiasm to apply our design method in multiple professional software for task efficiency. This opens the door to using scaffolded interfaces as a bridge between different applications and platforms.

Users could design their own scaffolded interfaces, built around the tasks and concepts they use most. Instead of learning a new interface for every software, they could keep a consistent interface setup that maps familiar controls to different software. For example, a character animator could keep the same rigging and keyframing scaffolded interface across software. This lowers the barrier to using the best software tool for each task. 
Such a system could also improve cross-application workflows. Many complex projects require switching between multiple applications and software. A shared scaffold layer based on common concepts could create a smoother, more unified experience. This can help reduce the mental effort required by switching contexts and learning new interfaces.

The main benefit can potentially mirror what we observed in our study: lower cognitive load and more focus on performing the task. Instead of mastering each software tool’s interface, users could invest time in learning fundamental concepts and refining their personal scaffolded interface. Building a universal system like this would require technical efforts to map workflows and concepts to the functionality of different underlying software for UI implementation. The potential to simplify cross-platform workflows makes it an interesting area for future research.

\section{Limitations and Future Work}
First, our current design uses user-selectable complexity levels for progressive tool disclosure. This is just one point on the spectrum from user-driven to system-driven control. Future research could explore system-driven interface adaptation based on performance metrics or user modeling, as well as mixed-initiative approaches where the system suggests level changes or adapts specific interface elements contextually. 

Second, some expert participants raised concerns about potential over-reliance on the scaffolded interface, which might limit learning of the full software. To address this, future designs could include cues like UI minimaps or short animations showing where tools come from. The scaffolded interface could also fade as users gain skill, encouraging further transition to the native UI.

Lastly, our implementation focused on building scaffolded interfaces in Blender for 3D modeling and animation tasks. Future research could test the implementation of scaffolded interfaces in other software and domains to explore broader use and impact.

\section{Conlusion}
Steep learning curves in professional software due to complex interfaces significantly hinder user adoption and skill development. Existing solutions often fail to provide integrated learning support within workflows. To address this, we introduce \papername{}, a method for designing scaffolded interfaces in professional software. Specifically, scaffolded interfaces are designed to present task-relevant tools, manage complexity through adjustable levels, organize the UI around workflow stages with integrated domain concepts, and connect users to native software interactions to support skill transfer. Our evaluation involving both beginners and experts showed that the implemented scaffolded interfaces in Blender significantly improved task performance for both user groups and supported learning for beginners. These findings demonstrate the effectiveness of combining task-awareness, conceptual organization, and progressive disclosure in interface design. We also provide insights for future research aimed at enhancing software learnability and user empowerment through scaffolded interfaces.


\bibliographystyle{IEEEtran}
\bibliography{bibliography}

\onecolumn

\section*{Appendix}

\subsection{LLM Prompt} \label{sec:llm_prompt}
\subsubsection{Workflow Analysis and Conceptual Decomposition} 
\begin{llmscriptblock}[title=]

You are an expert Blender user and workflow analyst specializing in task decomposition for UI design. Your objective is to analyze the specified Blender user task and break it down into its primary conceptual stages.
\vspace{1em}

\noindent\textbf{TASK TO ANALYZE:}
\par\noindent
\llmplaceholder{USER TASK DESCRIPTION}
\vspace{1em}

\noindent\textbf{METHOD:}
\par\noindent
Decompose the task into logical stages that represent a typical, efficient workflow a user would follow within Blender. Focus on the major distinct phases, moving from initial setup/preparation towards finalization/review. Aim for a reasonable number of stages that capture the core workflow without excessive granularity.
\vspace{1em}

\noindent\textbf{OUTPUT REQUIREMENTS:}
\par\noindent
Provide the output as a numbered list. Each list item must represent a distinct conceptual stage and include:
A concise \textbf{Stage Name}.
A brief, one-sentence \textbf{Description} clarifying the purpose of the stage.
\vspace{0.5em}
The output must be structured clearly to serve as the foundational organization for sections within a custom Blender UI panel designed to guide users through this specific workflow.
\vspace{1em}

\noindent\textbf{EXAMPLE OUTPUT FORMAT:}
\par
\begin{itemize}[label=\textendash, wide, nosep, leftmargin=1.5em, topsep=2pt]
    \item \textbf{Stage Name 1}: Brief Description.
    \item \textbf{Stage Name 2}: Brief Description.
    \item \dots
    \item \textbf{Stage Name n}: Brief Description.
\end{itemize}

\end{llmscriptblock}

\subsubsection{Tool Selection and Functionality Mapping} 
\begin{llmscriptblock}[title=]

You are an expert Blender Python API developer (Blender version 3.6). Your task is to identify relevant Blender functionalities for a specific stage of a larger workflow, intended for inclusion in a custom UI panel.
\vspace{1em}

\noindent\textbf{GIVEN WORKFLOW STAGE:}
\begin{itemize}[label=\textendash, wide, nosep, leftmargin=2em]
    \item Stage Name: \llmplaceholder{STAGE NAME}
    \item Stage Description: \llmplaceholder{STAGE DESCRIPTION}
\end{itemize}
\vspace{1em}

\noindent\textbf{TASK:}
\par\noindent
Identify the core Blender Python operators (\texttt{bpy.ops.*}) and key properties (\texttt{bpy.types.Scene}, \texttt{context.object.data.*}, \texttt{context.tool\_settings.*}, \texttt{bpy.props.*} if defined in a \texttt{PropertyGroup}) most commonly used and essential for achieving the goals described in the provided workflow stage. Prioritize stable and frequently used API elements.
\vspace{1em}

\noindent\textbf{OUTPUT REQUIREMENTS:}
\par\noindent
For each stage, output a bulleted list of suggested functionalities. Each item in the list must include:
\begin{itemize}[label=\textendash, wide, nosep, leftmargin=1.5em, topsep=2pt]
    \item \textbf{Identifier}:  The full Python identifier.
    \item \textbf{Type}:  Specify if it's an `Operator' or `Property'.
    \item \textbf{Rationale}:  Briefly explain why this function is relevant to the given stage's description/goal.
    \item \textbf{Context}:  Note any critical context requirements for the function to be active or relevant.
    \item \textbf{Expertise}:  The suggested level (\texttt{BEGINNER}, \texttt{INTERMEDIATE}, or \texttt{EXPERT}) with a brief justification (e.g., \texttt{BEGINNER} - Core operation for the task, \texttt{INTERMEDIATE} - Requires understanding of parameters, \texttt{EXPERT} - Used for complex optimization or non-standard workflows).
\end{itemize}
\vspace{1em}

\noindent\textbf{EXAMPLE OUTPUT FORMAT:}
\par\noindent 
\begin{itemize}[label=\textendash, wide, nosep, leftmargin=1.5em, topsep=2pt]
    \item \textbf{Identifier}: \texttt{bpy.ops.uv.unwrap}
    \item \textbf{Type}: \texttt{Operator}
    \item \textbf{Rationale}: Performs the primary UV unwrapping calculation based on marked seams.
    \item \textbf{Context}: Requires Edit Mode, faces selected. Usually invoked via the `U' key menu.
    \item \textbf{Expertise}: \texttt{BEGINNER} - Fundamental operation after marking seams.
\end{itemize}

\end{llmscriptblock}

\subsubsection{Code Generation for Scaffolded Interface Implementation}
\begin{llmscriptblock}[title=]

You are an expert Blender Python UI developer using the \texttt{bpy} API (Blender version 3.6). Generate Python code for a Blender UI Panel (\texttt{bpy.types.Panel}) based precisely on the following specifications.
\vspace{1em}

\noindent\textbf{SPECIFICATIONS:}

\begin{enumerate}[label=\arabic*., wide, topsep=0.5em, itemsep=0.5em]
    \item \textbf{Panel Configuration:}
        \begin{itemize}[label=\textendash, wide, nosep, leftmargin=1.5em, topsep=2pt]
            \item bl\_idname:  \llmplaceholder{USER\_PANEL\_IDNAME}
            \item bl\_label:  \llmplaceholder{USER\_PANEL\_LABEL}
            \item bl\_space\_type:  \texttt{VIEW\_3D}
            \item bl\_region\_type:  \texttt{UI}
            \item bl\_category:  \llmplaceholder{USER\_CATEGORY\_NAME}
            \item poll(cls, context):  \textnormal{Include a basic \texttt{poll(cls, context)} method checking for relevant context.}
        \end{itemize}

    \item \textbf{Settings Storage (PropertyGroup):}
        \begin{itemize}[label=\textendash, wide, nosep, leftmargin=1.5em, topsep=2pt]
            \item Define a \texttt{PropertyGroup} named \llmplaceholder{USER\_PROPGROUP\_NAME}.
            \item Attach this \texttt{PropertyGroup} to the scene via a \texttt{PointerProperty} named \llmplaceholder{USER\_PROP\_POINTER\_NAME}.
            \item The \texttt{PropertyGroup} contains an \texttt{EnumProperty} named \texttt{expertise\_level} with items (\texttt{'BASIC'}, \texttt{'INTERMEDIATE'}, \texttt{'EXPERT'}).
        \end{itemize}

    \item \textbf{Panel \texttt{draw()} Method Implementation:}
        \begin{itemize}[label=\textendash, wide, nosep, leftmargin=1.5em, topsep=2pt, itemsep=2pt]
            \item Retrieve the \texttt{PropertyGroup} instance via \texttt{context.scene.}\llmplaceholder{USER\_PROP\_POINTER\_NAME}. 
            \item Display a UI element (e.g., \texttt{layout.prop}) allowing the user to select the \texttt{expertise\_level}.
            \item Use the following list of conceptual stages to structure the panel: \llmplaceholder{LIST\_OF\_STAGE\_NAMES}.
            \item For each stage name in the list, create a distinct visual section using \texttt{layout.box()} labeled with the stage name.
            \item Within each stage's box, display the tools specified in the \llmplaceholder{TOOL\_MAPPING\_DICTIONARY} provided below.
            \item The \llmplaceholder{TOOL\_MAPPING\_DICTIONARY} maps stage names to a list of tool tuples. Each tuple contains: \texttt{(\textquotesingle API\_CALL\_STRING\textquotesingle, \textquotesingle UI\_LABEL\textquotesingle, ['LEVEL1', 'LEVEL2', \dots])}. \texttt{API\_CALL\_STRING} is the operator ID or property path. \texttt{UI\_LABEL} is the text for the button/property. The list contains the expertise levels for which the tool should be VISIBLE.
            \item Implement the visibility logic: only draw a tool if the current \texttt{expertise\_level} matches one of the levels in the tool's visibility list.
            \item Use \texttt{layout.operator()} for operators specified in \texttt{API\_CALL\_STRING}.
            \item Use \texttt{layout.prop(props\_instance, property\_name)} for properties specified in \texttt{API\_CALL\_STRING}.
            \item Add a \texttt{layout.separator()} between visible tools within each stage's box for visual spacing.
        \end{itemize}

    \item \textbf{Code Requirements:}
        \begin{itemize}[label=\textendash, wide, nosep, leftmargin=1.5em, topsep=2pt]
            \item Include necessary imports (\texttt{bpy}, \texttt{bpy.props}, \texttt{bpy.types}).
            \item Adhere to standard Blender Python scripting conventions and formatting.
            \item Include complete, basic registration and unregistration functions for the Panel class.
        \end{itemize}

    \item \textbf{Reference Code Structure:}
        \begin{itemize}[label=\textendash, wide, nosep, leftmargin=1.5em, topsep=2pt]
            \item You may refer to the provided \llmplaceholder{CODE\_EXAMPLE\_PYTHON\_SCRIPT} primarily to understand the expected structure of the \texttt{draw} method, conditional visibility checks based on expertise level, and the usage of \texttt{layout} elements (\texttt{box}, \texttt{prop}, \texttt{operator}, \texttt{separator}). Adapt the structure as needed to fit the specific tools and stages provided.
        \end{itemize}
\end{enumerate}
\vspace{1em}

\noindent\textbf{INPUTS TO BE USED:}
\begin{itemize}[label=\textendash, wide, nosep, leftmargin=2em, topsep=2pt]
    \item Conceptual Stages List: \llmplaceholder{LIST\_OF\_STAGE\_NAMES}
    \item Tool Mapping Dictionary: \llmplaceholder{TOOL\_MAPPING\_DICTIONARY}
    \item Code Example: \llmplaceholder{CODE\_EXAMPLE\_PYTHON\_SCRIPT}
\end{itemize}

\end{llmscriptblock}

\newpage
\subsection{User Study Materials}
\subsubsection{Study Interface} \label{sec:task2_interface}
\begin{figure*}[!ht]
    \centering
    \includegraphics[width=\textwidth]{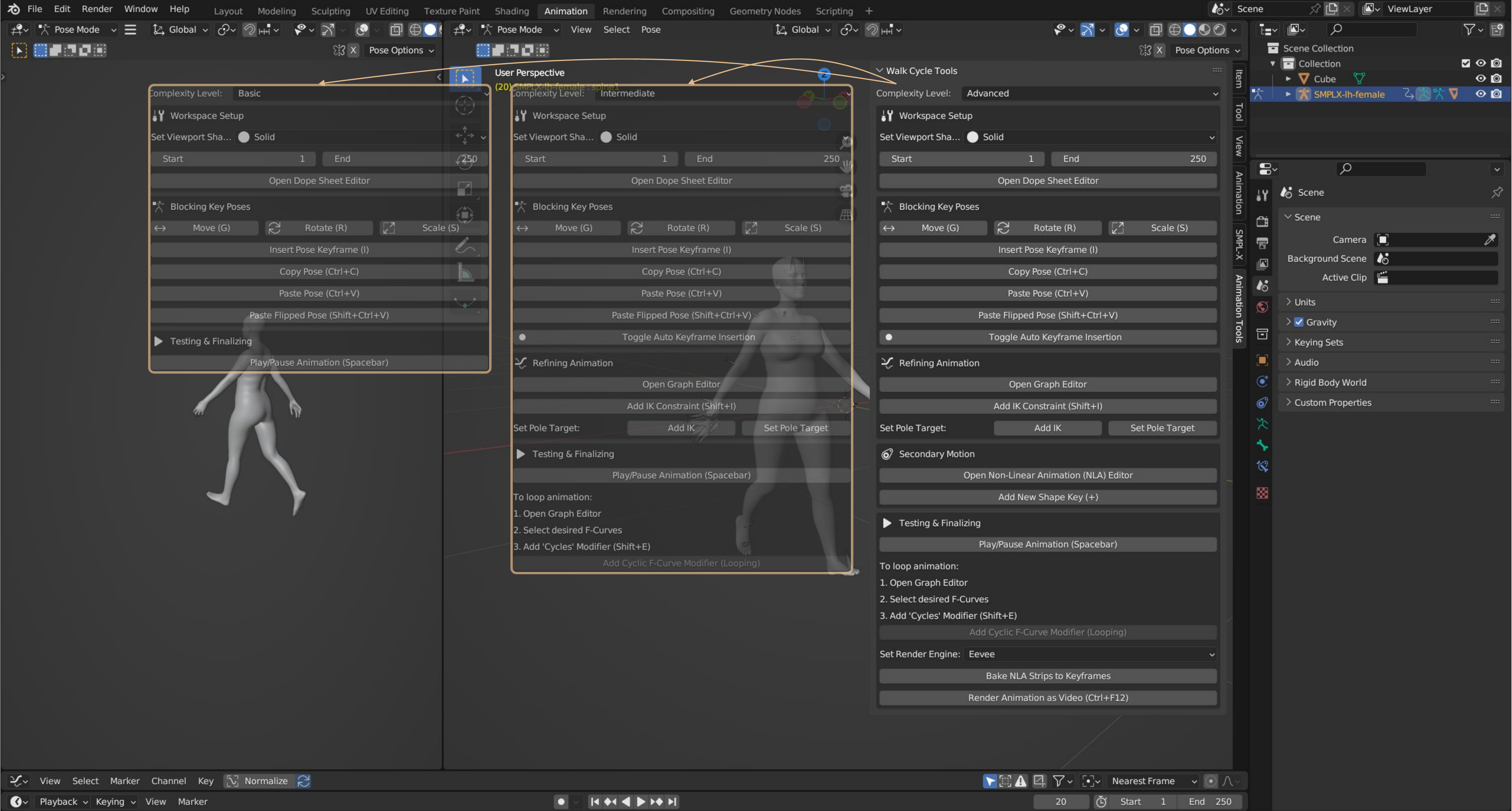}
    \caption{\papername{} implemented as a custom panel in Blender for the building walk cycle task. Task-relevant tools are grouped into logical sections that align with workflow stages and incorporate domain concepts. Each tool includes a descriptive label and a tooltip explaining its function and how it relates to domain concepts. To help users connect the scaffolded interface with Blender’s native UI, the interface displays keyboard shortcuts and hints about where each tool can be found in Blender’s standard menus or toolbars. The interface manages complexity via user-selectable levels for progressive tool disclosure.}
    \label{fig:task2_interface}
\end{figure*}

Figure \ref{fig:task2_interface} shows the user interface for Task 2: building a walk cycle.

\subsubsection{Study Guide} \label{sec:study_guide}
The video tutorial for UV unwrapping is available at \url{https://www.youtube.com/watch?v=7JUNlj6mR0U}, and for building a walk cycle can be found at \url{https://www.youtube.com/watch?v=e_COc0ZVHr0}.

\subsubsection{Questionnaires} \label{sec:study1_questionnaires}
Table \ref{tab:cognitive_load_questionnaire} introduces the questions we used from NASA-TLX to evaluate task load. The custom questionnaires for task performance, concept learning, and user interface experience are presented in Tables \ref{tab:task_performance_questionnaire}, \ref{tab:concept_learning_questionnaire}, and \ref{tab:interface_experience_questionnaire}. 

\begin{table}[!ht]
    \centering
    \caption{Task load questionnaire (NASA-TLX).}
    \scalebox{0.95} {
    \begin{tabular}{p{0.2\textwidth}p{0.5\textwidth}}
    \toprule
        Mental demand & How mentally demanding was it using the interface to perform your task? \\ 
        \midrule
        Temporal demand & How hurried or rushed was it using the interface to perform your task? \\ 
        \midrule
        Performance & How successful were you in accomplishing what you were asked to do? \\ 
        \midrule
        Effort & How hard did you have to work to accomplish your level of performance? \\ 
        \midrule
        Frustration & How insecure, discouraged, irritated, stressed, and annoyed were you? \\
    \bottomrule
    \end{tabular}
    }
    \label{tab:cognitive_load_questionnaire}
\end{table}

\begin{table}[!ht]
    \centering
    \caption{Task performance questionnaire.}
    \scalebox{0.95} {
    \begin{tabular}{p{0.2\textwidth}p{0.5\textwidth}}
    \toprule
        Workflow clarity & I felt clear about the task workflow. \\ 
        \midrule
        Task efficiency & I was able to complete the task efficiently. \\ 
        \midrule
        Tool relevance & It was easy for me to find the tools I needed to perform my task. \\ 
        \midrule
        User focus \& engagement & It was easy for me to stay focused on and engaged in my task. \\ 
        \midrule
        User confidence \& satisfaction & I felt confident performing the task and was satisfied with my accomplishment. \\
    \bottomrule
    \end{tabular}
    }
    \label{tab:task_performance_questionnaire}
\end{table}

\begin{table}[!ht]
    \centering
    \caption{Concept learning questionnaire.}
    \scalebox{0.95} {
    \begin{tabular}{p{0.2\textwidth}p{0.5\textwidth}}
    \toprule
        Concept-workflow correlation & I was able to correlate task-related concepts with the task workflow. \\ 
        \midrule
        Concept understanding & Using this interface helped me to understand these task-related concepts. \\ 
        \midrule
        Concept-tool correlation & I was able to correlate tools used for my task with these task-related concepts. \\ 
        \midrule
        Learning motivation & Using this interface enhanced my motivation and experience of learning concepts. \\ 
        \midrule
        Adjacent concept exploration & I was able to explore concepts beyond those directly related to my task. \\
    \bottomrule
    \end{tabular}
    }
    \label{tab:concept_learning_questionnaire}
\end{table}

\begin{table}[!ht]
    \centering
    \caption{User interface experience questionnaire.}
    \scalebox{0.95} {
    \begin{tabular}{p{0.2\textwidth}p{0.5\textwidth}}
    \toprule
        Tool support vs. distraction & The interface offered relevant tools that supported my task without unnecessary distractions. \\ 
        \midrule
        Prior knowledge requirement & Prior interface familiarity (e.g., tool locations, usage, keyboard shortcuts) was not required to perform my task. \\ 
        \midrule
        External help requirement & Extra assistance (e.g., referring to tutorials or asking questions) was not required to perform my task. \\ 
        \midrule
        Adjacent tool exploration & I was able to explore tools in the interface beyond those directly related to my task. \\ 
        \midrule
        Interface vs. user adaptation & The interface appeared tailored to support my task rather than requiring me to adapt to it. \\
    \bottomrule
    \end{tabular}
    }
    \label{tab:interface_experience_questionnaire}
\end{table}

\subsubsection{Open-Ended Questions} \label{sec:study2_openended_questions}
\textit{Compared to the default Blender interface, please provide your experience with the custom interface that organizes tools according to your task workflow and task-related concepts, and supports progressive disclosure of tools. We would appreciate detailed responses.}

1. Did the custom interface help you complete the task better or worse than you expected? Were you able to easily find what you needed in the interface to perform your task? How did it make you feel while working - for example, did you feel engaged or sometimes lost? Did the layout and design of the interface make sense to you?

2. (Task 1: UV Unwrapping) If you were to learn new concepts, would the custom interface help you understand the concepts better or make them more confusing? For example, would things like seams, unwrapping, and UV stretching be easier to grasp? Would it help you discover anything new that isn’t directly part of your task but a related idea such as UV islands? Would the way the interface presents the individual tools make learning feel natural or difficult?

2. (Task 2: Building Walk Cycle) If you were to learn new concepts, would the custom interface help you understand the concepts better or make them more confusing? For example, would things like key poses and bone constraints be easier to grasp? Would it help you discover anything new that isn’t directly part of your task but a related idea such as secondary motion? Would the way the interface presents the individual tools make learning feel natural or difficult?

3. How did the custom interface affect your overall experience? Did it feel like it was supporting you or getting in your way? Could you figure things out on your own, or did you need help? Did you feel like you had to adjust to the system, or did it adjust to you and your needs?

4. If you were to use the custom interface for a longer period, would it make your long-term skill development easier or harder? For example, would building up your skills from basic to advanced be easier, and if so, how? Would it help you build practical skills by teaching you underlying concepts? Would the learned concepts be transferable?

5. If you have experience with other software similar to Blender (e.g., Maya, 3ds Max), could the methodology of clustering related tools and gradually disclosing them as users make progress used to build the custom interface be applied to build similar interfaces for them? If yes, how might it be used, and what potential impact could it have?

6. Are there any features you would add, remove, or modify in the custom interface? If yes, please specify.

\subsection{User Study 2 Participant Demographic Information} \label{sec:study2_demographics}
Table \ref{tab:study2_demographics} presents the demographic information of Study 2 participants. 

\begin{table*}[!ht]
    \centering
    \caption{Study 2 participant demographic information.}
    \scalebox{0.95} {
    \begin{tabular}{ccccl}
    \toprule
        ID & Year of Experience & 3D Modeling & Animation & \multicolumn{1}{c}{Additional Software Experience} \\
        \midrule
        S2-P1 & 2 & \cmark & \xmark & SketchUp \\
        \midrule
        S2-P2 & 5 & \cmark & \cmark & Maya, ZBrush, MotionBuilder \\
        \midrule
        S2-P3 & 3 & \cmark & \cmark & Maya \\
        \midrule
        S2-P4 & 3 & \cmark & \cmark & 3ds Max, Mixamo, Unity, Unreal Engine \\
        \midrule
        S2-P5 & 6 & \cmark & \cmark & Unity, Maya \\
        \midrule
        S2-P6 & 4 & \cmark & \cmark & Tinkercad, 3ds Max, Maya \\
        \midrule
        S2-P7 & 1 & \cmark & \xmark & FreeCAD, Tinkercad \\
        \midrule
        S2-P8 & 5 & \cmark & \cmark & Substance 3D, After Effects \\
    \bottomrule
    \end{tabular}
    }
    \label{tab:study2_demographics}
\end{table*}

\subsection{User Study 1 ANOVA Results} \label{sec:study1_anova}
Tables \ref{tab:anova_taskload} to \ref{tab:anova_interfaceexperience} report the two-way ANOVA results of Study 1.

\begin{table}[!ht]
    \centering
    \caption{Two-way ANOVA results for task load measures.}
    \label{tab:anova_taskload}
    \scalebox{0.95} {
    \begin{tabular}{p{0.2\textwidth} cc cc cc} 
    \toprule
    & \multicolumn{2}{c}{Interface Effect} & \multicolumn{2}{c}{Task Effect} & \multicolumn{2}{c}{Interaction Effect} \\
    \cmidrule(lr){2-3} \cmidrule(lr){4-5} \cmidrule(lr){6-7} 
                    & $F_{(1, 30)}$ & $p$     & $F_{(1, 30)}$ & $p$     & $F_{(1, 30)}$ & $p$     \\ 
    \midrule
    Mental demand   & $22.53$  & $ < .001 $ & $8.53$  & $.007$    & $0.13$  & $.718$    \\
    Temporal demand & $95.38$  & $ < .001 $ & $4.71$  & $.039$    & $0.52$  & $.475$    \\
    Performance     & $150.36$ & $ < .001 $ & $0.19$  & $.665$    & $1.73$  & $.200$    \\
    Effort          & $69.96$  & $ < .001 $ & $8.51$  & $.007$    & $8.51$  & $.007$    \\
    Frustration     & $91.43$  & $ < .001 $ & $4.32$  & $.047$    & $2.77$  & $.107$    \\
    \bottomrule
    \end{tabular}
    }
\end{table}

\begin{table}[!ht]
    \centering
    \caption{Two-way ANOVA results for task performance measures.}
    \label{tab:anova_taskperformance} 
    \scalebox{0.95} { 
    \begin{tabular}{p{0.2\textwidth} cc cc cc}
    \toprule
    & \multicolumn{2}{c}{Interface Effect} & \multicolumn{2}{c}{Task Effect} & \multicolumn{2}{c}{Interaction Effect} \\
    \cmidrule(lr){2-3} \cmidrule(lr){4-5} \cmidrule(lr){6-7}
                            & $F_{(1, 30)}$ & $p$     & $F_{(1, 30)}$ & $p$     & $F_{(1, 30)}$ & $p$     \\
    \midrule
    Workflow clarity        & $34.60$  & $ < .001 $ & $2.05$  & $.164$    & $0.63$  & $.433$    \\
    Task efficiency         & $19.96$  & $ < .001 $ & $12.55$ & $.001$    & $0.21$  & $.648$    \\
    Tool relevance          & $48.13$  & $ < .001 $ & $0.53$  & $.471$    & $0.13$  & $.718$    \\
    User focus \& engagement        & $71.46$  & $ < .001 $ & $0.42$  & $.521$    & $2.30$  & $.140$    \\
    User confidence \& satisfaction & $38.53$  & $ < .001 $ & $2.13$  & $.155$    & $4.80$  & $.037$    \\
    \bottomrule
    \end{tabular}
    }
\end{table}

\begin{table}[!ht]
    \centering
    \caption{Two-way ANOVA results for concept learning measures.}
    \label{tab:anova_conceptlearning}
    \scalebox{0.95} { 
    \begin{tabular}{p{0.2\textwidth} cc cc cc}
    \toprule
    & \multicolumn{2}{c}{Interface Effect} & \multicolumn{2}{c}{Task Effect} & \multicolumn{2}{c}{Interaction Effect} \\
    \cmidrule(lr){2-3} \cmidrule(lr){4-5} \cmidrule(lr){6-7}
                            & $F_{(1, 30)}$ & $p$     & $F_{(1, 30)}$ & $p$     & $F_{(1, 30)}$ & $p$     \\
    \midrule
    Concept-workflow correlation  & $78.91$  & $ < .001 $ & $7.39$  & $.011$    & $0.03$  & $.857$    \\
    Concept ynderstanding   & $29.67$  & $ < .001 $ & $2.93$  & $.098$    & $0.02$  & $.877$    \\
    Concept-tool correlation      & $33.39$  & $ < .001 $ & $1.98$  & $.171$    & $1.20$  & $.284$    \\
    Learning motivation     & $36.25$  & $ < .001 $ & $0.49$  & $.490$    & $0.18$  & $.678$    \\
    Adjacent concept exploration  & $41.20$  & $ < .001 $ & $10.30$ & $.003$    & $0.16$  & $.691$    \\
    \bottomrule
    \end{tabular}
    }
\end{table}

\begin{table}[!ht]
    \centering
    \caption{Two-way ANOVA results for user interface experience measures.}
    \label{tab:anova_interfaceexperience} 
    \scalebox{0.95} { 
    \begin{tabular}{p{0.2\textwidth} cc cc cc}
    \toprule
    & \multicolumn{2}{c}{Interface Effect} & \multicolumn{2}{c}{Task Effect} & \multicolumn{2}{c}{Interaction Effect} \\
    \cmidrule(lr){2-3} \cmidrule(lr){4-5} \cmidrule(lr){6-7}
                            & $F_{(1, 30)}$ & $p$     & $F_{(1, 30)}$ & $p$     & $F_{(1, 30)}$ & $p$     \\
    \midrule
    Tool support vs. distraction        & $44.44$  & $ < .001 $ & $0.44$  & $.510$    & $0.00$  & $1.000$   \\
    Prior knowledge requirement    & $32.36$  & $ < .001 $ & $11.65$ & $.002$    & $2.14$  & $.155$    \\
    External help requirement      & $26.75$  & $ < .001 $ & $15.35$ & $.001$    & $4.15$  & $.051$    \\
    Adjacent tool exploration     & $59.25$  & $ < .001 $ & $3.25$  & $.082$    & $0.67$  & $.420$    \\
    Interface vs. user adaptation         & $104.14$ & $ < .001 $ & $0.14$  & $.708$    & $1.29$  & $.266$    \\
    \bottomrule
    \end{tabular}
    }
\end{table}

\clearpage
\subsection{Scaffolded Interface Variations} \label{sec:interface_variations}
\subsubsection{UV Unwrapping}
Figure \ref{fig:task1_interface_vars} shows the scaffolded interface variations for UV unwrapping. 

For \textit{organic} models like characters and creatures, the workflow focuses on creating smooth, low-distortion texture maps. The interface helps users define seams to flatten curved surfaces, apply algorithms that preserve surface flow, and refine UV layouts to keep textures accurate during animation.

For \textit{hard-surface} models, the workflow is designed to handle their unique geometry and texturing needs. The interface supports multiple UV sets for different textures, uses precise projection and alignment tools to create clean UV islands, and includes tools for accurate decal placement.

\subsubsection{Building Walk Cycle}
\begin{figure*}[!ht]
    \centering
    \subfloat[UV Unwrapping]{
        \centering
        \includegraphics[width=0.48\textwidth]{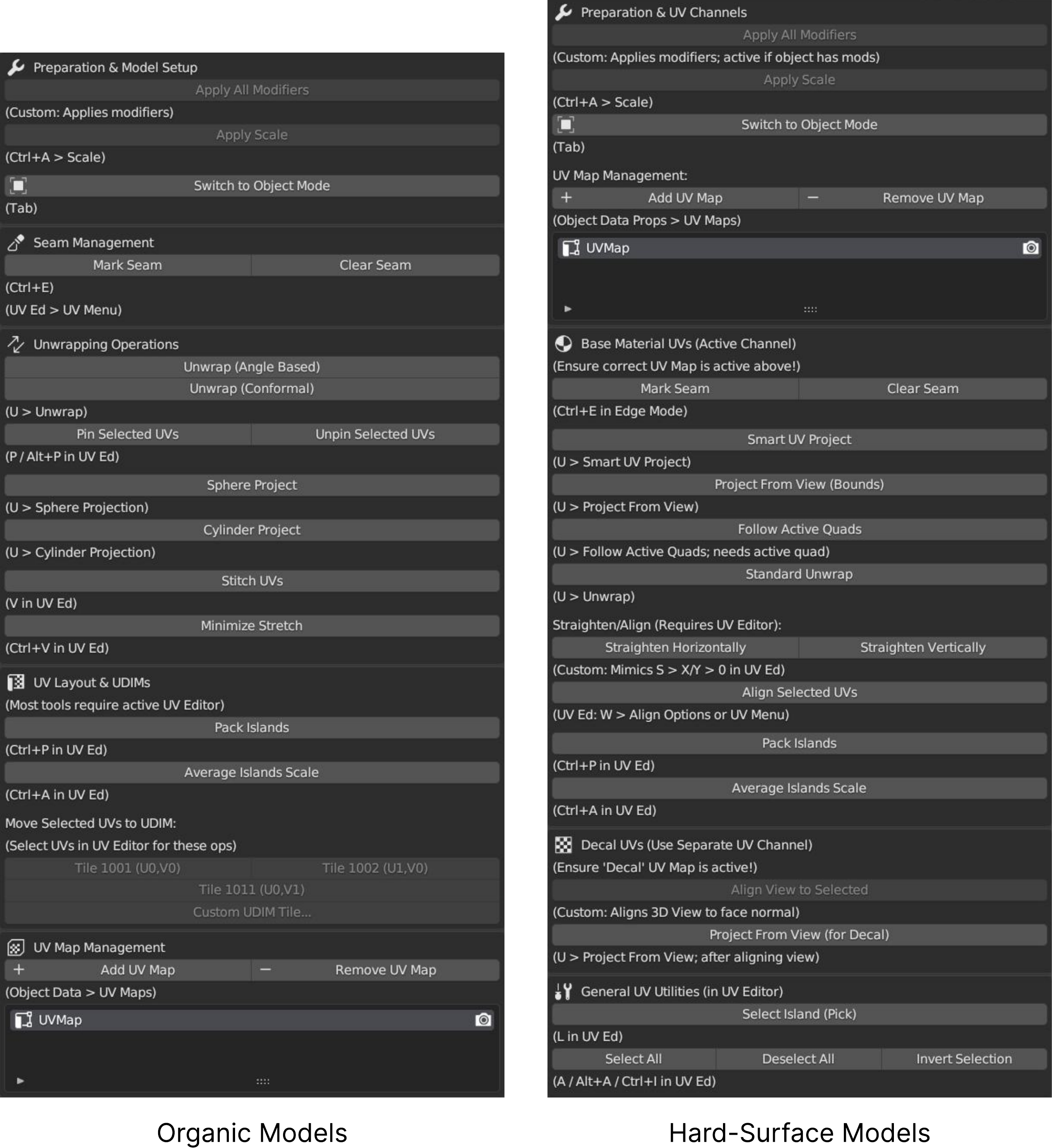}
        \label{fig:task1_interface_vars}
    }
    \hfill
    \subfloat[Building Walk Cycle]{
        \centering
        \includegraphics[width=0.48\textwidth]{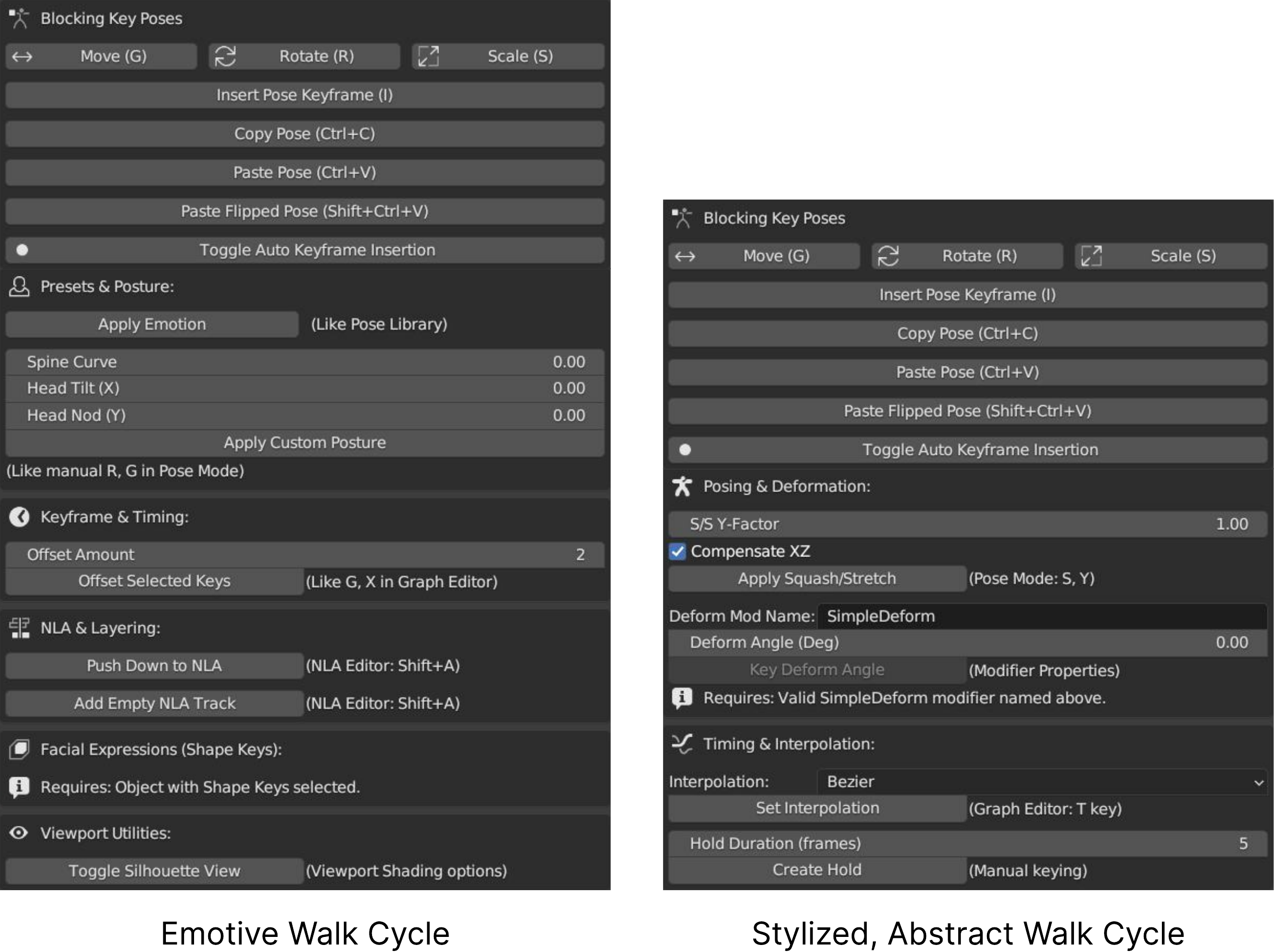}
        \label{fig:task2_interface_vars}
    }
    \caption{\papername{} interface variations.}
    \label{fig:interface_vars}
\end{figure*}

Figure \ref{fig:task2_interface_vars} presents the scaffolded interface variations for building a walk cycle. 

The \textit{emotive walk cycle} workflow adds emotion and personality to walk animations. The interface guides users in analyzing reference material, adjusting key poses, especially spine and head, and refining timing and curves, with layered secondary motion to express the desired mood or personality.

The \textit{stylized, abstract walk cycle} workflow focuses on creating animations that move away from realism for artistic effect. The interface supports exaggerating form and motion, adjusting timing and interpolation for unique rhythms, and creating bold, non-realistic effects to match artistic or conceptual goals.

\end{document}